\documentclass[fleqn,usenatbib]{mnras}
\usepackage{newtxtext,newtxmath}

\usepackage[T1]{fontenc}
\usepackage{ae,aecompl}
\usepackage{graphicx,epsfig}	
\usepackage{amsmath}	
\usepackage{array}
\usepackage{booktabs} 
\usepackage[dvipsnames]{xcolor}
\usepackage{url}
\newcommand{\re}{R_{\rm e}}
\newcommand{\msun}{M$_\odot$}

\newcommand{\mstar}{M$_{*}$}

\newcommand{\kms}{km~s${-1}$}
\newcommand{\un}{$\,\pm\,$}
\newcommand{\ssc}{S\'{e}rsic}

\title[The realm of cE galaxies with KCWI]{Low-mass compact elliptical galaxies: spatially-resolved stellar populations and kinematics with the Keck Cosmic Web Imager}

\author[A. Ferr\'e-Mateu et  al.]{Anna Ferr\'e-Mateu$^{1,2}$\thanks{E-mail: aferremateu@icc.ub.edu (AFM)}, Mark Durre$^{2}$, Duncan A.\ Forbes$^{2}$, Aaron J.\ Romanowsky$^{3,4}$,
\newauthor
Adebusola Alabi$^{3}$, Jean P.\ Brodie$^{2,3}$ and Richard M. McDermid$^{5,6}$\\
\\
$^{1}$ Institut de Ciencies del Cosmos (ICCUB), Universitat de Barcelona (IEEC-UB), E02028 Barcelona, Spain\\
$^{2}$ Centre for Astrophysics \& Supercomputing, Swinburne University of Technology, Hawthorn VIC 3122, Australia\\
$^{3}$ University of California Observatories, 1156 High St., Santa Cruz, CA 95064, USA\\
$^{4}$ Department of Physics and Astronomy, San Jos\'e State University, San Jose, CA 95192, USA \\
$^{5}$ Department of Physics and Astronomy, Macquarie University, Sydney NSW 2109, Australia\\
$^{6}$ARC Centre of Excellence for All Sky Astrophysics in 3 Dimensions (ASTRO 3D)\\
}

\date{Accepted 2021 March 11. Received 2021 February 16; in original form 2020 December 10}
\pubyear{2021}

\begin{document}
\label{firstpage}
\pagerange{\pageref{firstpage}--\pageref{lastpage}}
\maketitle

\begin{abstract}
We present spatially-resolved two-dimensional maps and radial trends of the stellar populations and kinematics for a sample of six compact elliptical galaxies (cE) using spectroscopy from the Keck Cosmic Web Imager (KCWI). We recover their star formation histories, finding that all except one of our cEs are old and metal rich, with both age and metallicity decreasing toward their outer radii. We also use the integrated values within one effective radius to study different scaling relations. Comparing our cEs with others from the literature and from simulations we reveal the formation channel that these galaxies might have followed. All our cEs are fast rotators, with relatively high rotation values given their low ellipticites.  In general, the properties of our cEs are very similar to those seen in the cores of more massive galaxies, and in particular, to massive compact galaxies. Five out of our six cEs are the result of stripping a more massive (compact or extended) galaxy, and only one cE is compatible with having been formed intrinsically as the low-mass, compact object that we see today. These results further confirm that cEs are a mixed-bag of galaxies that can be formed following different formation channels, reporting for the first time an evolutionary link within the realm of compact galaxies (at all stellar masses).  
\end{abstract}

\begin{keywords}
galaxies: evolution -- galaxies: formation -- galaxies: kinematics and dynamics -- galaxies: stellar content 
\end{keywords}

\section{Introduction}

The low-mass and compact regime of early-type galaxies (ETGs) is populated by different families of galaxies (e.g. \citealt{Drinkwater2000}; \citealt{Hasegan2005}; \citealt{Brodie2011}; \citealt{Misgeld2011}). One of these is the so-called compact ellipticals (cE), broadly characterized by galaxies having stellar masses of 10$^{8}$ \la M$_{*}$/M$_{\odot} \la $10$^{10}$ and galaxy sizes between 100 -- 900\,pc. This implies high stellar densities that resemble those in the cores of massive ETGs or the bulges of spirals (e.g.\ \citealt{Faber1973}; \citealt{Bekki2001}; \citealt{Choi2002}; \citealt{Graham2002}; \citealt{Drinkwater2003}). However, their origins and relationship to the more massive and extended ETGs are still under debate, with different formation channels proposed. 

One possibility is that cEs are formed by stripping the loosely-bound stars from the outer envelopes of a larger, more massive galaxy (i.e.\ \textit{nurture}). In this case, cEs should reveal the properties of the central regions of the progenitor. The irrefutable smoking gun for this scenario is the fact that some cEs have been seen being stripped by their host galaxy (e.g.\ \citealt{Huxor2011}; \citealt{Paudel2014}). Moreover, some of them have been found to host larger super-massive black holes (SMBH) than expected for their stellar mass (e.g.\ \citealt{vanderMarel1997};  \citealt{Barber2016}; \citealt{Paudel2016}; \citealt{Pechetti2017}). This can be easily explained under the stripping origin, as the measured SMBH would correspond to the one formed initially with the massive progenitor, before it was stripped (e.g. \citealt{Ferre-Mateu2015}; \citealt{vanSon2019}). 

However, some cEs have also been discovered in isolation, with no possible host nearby to produce the stripping, and thus an intrinsic, or in-situ, origin has been proposed for these (e.g.\ \citealt{Huxor2013}; \citealt{Paudel2014b}). In this scenario, such cEs are formed as intrinsically compact objects as we see them today (i.e.\ \textit{nature}), representing the lowest mass/luminosity end of the classical ETG family (e.g.\ \citealt{Wirth1984}; \citealt{Nieto1987}; \citealt{Kormendy2009}; \citealt{Kormendy2012}). As such, these cEs are expected to follow the scaling relations that govern classical ETGs. For example, Under this assumption they should host the elusive and long sought intermediate black holes (\citealt{Mezcua2017}) that correspond to galaxies of such low stellar masses. Nonetheless, there is also a more exotic nurture channel proposed for these isolated cEs, wherein they have been ejected far from their massive hosts through a three-body encounter, after tidal stripping \citep{Chilingarian2015}.

A handful of simulations aim to reproduce the different formation channels of cEs, owing to the different origins. \citet{Martinovic2017} simulated cEs located near ($<$100 kpc) a massive galaxy, and found that the majority formed intrinsically as a low mass compact object, in-situ within a cluster environment. Such a dense environment was responsible for preventing the galaxy growing further in size (e.g.\ \citealt{Wellons2016}). Only 30\% of their simulated cEs were formed outside the cluster as larger, Milky-Way-type galaxies which subsequently had their stars stripped through cluster infall. Exploring a different channel, \citet{Du2018} used high resolution simulations to reproduce cEs as the product of a low-mass satellite galaxy infalling on highly radial orbits into a more massive galaxy. They found that tidal stripping from the massive galaxy alone is not enough to produce a cE, and instead, a combination of ram-pressure confinement, tidal stripping and bursty star formation is required. Nonetheless, these two simulations were tailored for dense environments and therefore can not explain the existence of cEs in the field. To cover this, \citet{UrrutiaZapata2019} investigated another mechanism for forming cEs. They simulated their cEs as the result of several mergers of very massive star clusters \citep{Fellhauer2002}, producing the most compact and least massive cEs seen observationally. Therefore, three different sets of simulations can already reproduce the properties of observed cEs via different formation channels, but none alone is enough to describe the entire family, reinforcing the view that a variety of origins exist for this rare family.

\begin{table*}
\centering
\caption{Summary of the targets and observational parameters.}
\label{tab:obs}
\begin{tabular}{@{}lccllcccc@{}}
\toprule
Galaxy              & RA          & Dec            & Host     & Proj.D  & Distance  & Source         & Configuration & Exposures (s)     \\ 
(1)                 & (2)         & (3)            & (4)      &  (5)          & (6)       & (7)            & (8)           &  (9)    \\
\midrule
J160537.21+142441.2 & 16:05:37.2  & $+$14:24:41  &  --      &  None     & 80.8\un5.7 & a & Small  BL4550 & 2 $\times$ 1800          \\
NGC~3665cO       & 11:24:54.7  & $+$38:46:44  &  NGC~3665 & 141.8  & 37.7\un2.6 & a & Large  BL4550 & 3 $\times$ 600           \\
NGC~4486B            & 12:30:31.9  & $+$12:29:24  &  NGC~4486 & 394.5  & 16.5\un2.3 & b & Large  BL4550 & 2 $\times$ 1200          \\
NGC~5846A            & 15:06:29.4  & $+$01:35:41  &  NGC~5846 &  38.7  & 25.0\un3.0 & b & Large  BL4550 & 3 $\times$ 600 + 6 $\times$ 400 \\
NGC~5846cE       & 15:06:34.2  & $+$01:33:31  &  NGC~5846 & 310.3  & 25.0\un3.0 & b & Medium BL4550 & 2 $\times$ 1200          \\
VCC~344              & 12:19:22.1  & $+$05:47:56  &  NGC~4261 & 108.8  & 32.4\un4.5 & b & Small  BL4550 & 4 $\times$ 900           \\ 
\bottomrule
\end{tabular}
\\
\vspace{+0.2cm}
Columns: (1) Galaxy; (2--3) coordinates in J2000, (4) host galaxy/group/cluster; (5) projected distance to the host (arcsec); \\
(6) distance (Mpc); 
(7) source for distance: (a) - NED, (b) - \cite{Tully2013}; \\
(8) KCWI configuration of slicer field of view, grating and central wavelength; (9) exposure times for each object. 
\end{table*}

Given the limited theoretical information and the rarity of these objects, they constitute an open puzzle for the low mass/luminosity end of the ETG realm. From the few couple hundreds reported to date (\citealt{Chilingarian2009}; \citealt{Norris2014}; \citealt{Chilingarian2015}; \hypertarget{K+20}{\citealt{Kim2020}}, K+20 hereafter), it is still unclear if there is any mechanism that dominates their formation and whether or not these are related to any environmental or mass dependencies. Fortunately, cEs show very distinctive trends in their stellar population relations that can help differentiate between one origin or another. For example, cEs with a stripped origin will be outliers in the local mass--metallicity relations (e.g. \citealt{Lequeux1979}; \citealt{Matteucci1994}; \citealt{Gallazzi2005}; \citealt{Kirby2013}), being more metal-rich than expected from their stellar mass. In contrast, those cEs that follow the mass--metallicity relation suggest an intrinsic origin. Following the idea of having several indicators to differentiate between origins, we analyzed in \hypertarget{FM+18}{\citet{Ferre-Mateu2018}} (FM+18 hereafter) the properties of a sample of 25 cEs. We found that about 85\% of the cEs were compatible with a stripping origin (previously or ongoing). The remaining 15\% were better represented by intrinsically low-mass compact systems. Interestingly, it turned out that these intrinsic cEs were mostly those with no nearby host and located in sparse environments. We note that this sample was biased towards objects near a host galaxy. Using a larger and more homogeneous sample of cEs, \hyperlink{K+20}{K+20} found that from their sample of 138 cEs, 65 could be associated to a host galaxy while 73 were either isolated or not close enough to a more massive galaxy. However, both the \hyperlink{FM+18}{FM+18} and \hyperlink{K+20}{K+20} works are based on SDSS fibre spectroscopy, whose fixed aperture size leads to a range of radial coverage depending on the distance of the galaxy, in these cases ranging from $0.5 \re$ to $3 \re$ (effective radii). This limitation can introduce biases to the derived mass--metallicity relation in the presence of metallicity gradients.

In this vein we now study the spatially-resolved 2D properties of this unique and rare family of galaxies to better understand their different formation pathways. Although cEs have high central densities, their surface brightness decline rapidly with radius. We are thus pushing the limits of the new IFU capabilities by obtaining spatially resolved properties of these galaxies and probing into their outskirts when possible. In this study, one of the very few to date, we present a sample of six cEs using the Keck Cosmic Web Imager (KCWI; \citealt{Morrissey2012}) on the Keck-II telescope. Most integral-field unit (IFU) studies in this low-mass/luminosity regime have focused on the more common, extended dEs (e.g.\ \citealt{Rys2014}; \citealt{Toloba2014}; \citealt{Scott2020}), and therefore the low-mass, compact regime of ETGs remains mostly unexplored. Only \hypertarget{G+15}{\citet{Guerou2015}} (G+15 hereafter) previously obtained 2D information for a sample of cEs, although their sample did not reach the lowest mass range defined for cEs. The sample presented in this work, despite being equally small, covers the entire mass range expected for cEs. We define this work sample in \S \ref{section:sample}, presenting the new KCWI observations and the data reduction in \S \ref{section:data}. \S \ref{section:ana} presents the analysis of both the stellar populations and kinematics from the resulting 2-D spectroscopic maps, \S \ref{section:results} discusses our results and the implications for the possible formation pathways of cEs and \S \ref{section:conc} presents the conclusions. 

\section{The sample}\label{section:sample}
While spectroscopic studies of cEs have been done with a few tens of candidates, only about a dozen objects have been studied in detail, mostly with long-slit spectroscopy (e.g.\  \citealt{Graham2002}; \citealt{Choi2002}; \citealt{Huxor2011}; \citealt{Huxor2013}; \citealt{Paudel2013}; \citealt{Janz2017}; \citealt{Paudel2014b}). In \hyperlink{FM+18}{FM+18} we studied the stellar populations and kinematics of a sample of 25 cEs from SDSS. However, the study was spatially-restricted to the fibre size of SDSS (i.e.\ $\sim$3" diameter), which meant that we only partially covered some of the cEs, introducing possible aperture effects. In fact, IFU observations of cE are extremely scarce due to the their small sizes and low luminosities, with the notable exception of \hyperlink{G+15}{G+15}. In the latter, the authors performed a kinematic and stellar population analysis for a sample of 8 cEs in Virgo with 
Gemini/GMOS. In this paper we aim at building up the sample of spatially-resolved cEs with detailed stellar populations and kinematic analysis in order to further understand this rather poorly-studied family of ETGs in the low mass regime. We are able to almost double the number of spatially-resolved cEs, extending the analysis to the lowest masses in the cE range and to other environments outside of the Virgo Cluster from \hyperlink{G+15}{G+15}.

Our sample of cE includes some already known ones, such as NGC~5846A, NGC~4486B (= VCC~1297), and VCC~344. It also includes other cEs that have been recently discovered, like NGC~5846cE and J160537.21+142441.2 (\hyperlink{FM+18}{FM+18}). The latter was renamed J1614 for simplicity and we keep here this naming. For J1614, NGC~4486B, NGC~5846cE and VCC~344, there are available photometric SDSS data, and NGC~4486B was also studied by \hyperlink{G+15}{G+15}. NGC~3665cO is a new discovery reported here for the first time. It was part of an ongoing search for cEs and high-mass ultracompact dwarfs (UCDs) around host galaxies from the ATLAS$^{\rm 3D}$ survey \citep{Cappellari2011}. Following our success of using SDSS {\tt SQL} searches for compact galaxies (\citealt{Sandoval2015}; \hyperlink{FM+18}{FM+18}) we used the DR7 database to search for objects within 40~kpc projected distance of a host galaxy, with photometric parameters appropriate to cEs and UCDs. The criteria included: SDSS photometric classification as ``GALAXY'' based on being non-pointlike; absolute magnitude in the range $M_r = -12.0$ to $-20.0$ (equivalent to $M_* \sim\,10^7$--$10^{10}$ M$_{\odot}$); colour $g-i = $~0.9--1.4; half-light radius R$_{50} =$~40~pc--1~kpc (note these SDSS model-based sizes are comparable to the seeing and should be treated as initial guesses only). The search returned compact objects like SDSS J112454.73+384644.4, close to NGC~3665. We decided to obtain IFU observations because this is the most compact, least massive of the cEs ever observed, being close to the limit on what can be considered a cE. For this reason we dub it NGC~3665cO (compact object) rather than NGC~3665cE. Objects like this are crucial to explore the limiting regions between types of galaxies. VCC~344, also resulted from this search. Established as part of the Virgo Cluster Catalogue (\citealt{Binggeli1985}; but actually it is behind the Virgo Cluster), it was called ``M32-like'' by \citet{Sandage1994} and classified as ``cE2'' in NED -- yet it was missed in works on nearby cEs in the recent years.

All of the galaxies in the sample except for J1614 are associated with a larger, more massive galaxy (see Table \ref{tab:obs}). J1614 is an interesting case because it resides in the outskirts of a galaxy cluster but there is no plausible host within a radius of several hundred kpc (see \hyperlink{FM+18}{FM+18} for more information). We note the first caveat of our sample, similar to \hyperlink{FM+18}{FM+18}; for each cE without host, there are 5 cEs with host. This selection bias disappears when larger and more homogeneous samples are used. For example, in \hyperlink{K+20}{K+20} $\sim$47\% of the cEs are associated to a host galaxy while the rest are not. Like \hyperlink{FM+18}{FM+18}, \hyperlink{K+20}{K+20} generally found that those with a host were located in clusters or groups, while the cEs without host were typically in the field. In this work, three cEs are in cluster environments while the other three belong to groups, but none is considered to be in the field. Nonetheless, one of the cluster cEs, J1614, is located in the outskirts of the cluster and could be considered as a low density region. Furthermore, deep imaging from the literature shows no indications that any of our targets are currently suffering an interaction with their host galaxy or surroundings, which means that they will be either already stripped galaxies or be of the intrinsic type. 

\section{Observations and data reduction}\label{section:data}
Observations of the six cEs were carried out during several nights in April and May 2018 (program IDs N79 and U250) using the KCWI integral field unit \citep{Morrissey2012} on the Keck II telescope. Most of the nights were clear with a seeing ranging from 0.6" to 1", except during the observations of NGC~3665cO, when the seeing was worse ($\sim$1.5") and there were some clouds. We proceed with this galaxy together with the rest, but we warn that its 2D maps should be taken with caution. The observations were taken with the BL grating centered at 4550\,\AA, with different slicers (as indicated in Table \ref{tab:obs}). These set-ups provide a good spectral range to perform stellar population and kinematic studies. The different slicers provide different spectral resolutions, which range from $\sim$5.5\,\AA\ ($\sigma \sim$ 140 km\,s$^\mathrm{-1}$) for the large slicer, to $\sim$2.7\,\AA\ ($\sigma \sim$ 71 km\,s$^\mathrm{-1}$) for the medium, and to $\sim$1.4\,\AA\ ($\sigma \sim$ 36 km\,s$^\mathrm{-1}$) for the small one. Table \ref{tab:obs} also shows the integration times for each object in our sample. Several standard stars were observed during the nights with the same set-ups, for flux calibration purposes. 

The data were reduced using the KCWI pipeline KDERP\footnote{\url{github.com/Keck-DataReductionPipelines/KcwiDRP}}, which performs a full standard data reduction. It delivers wavelength- and flux-calibrated 3D datacubes. The only stage we excluded from the pipeline was sky subtraction, which was performed as discussed below. Individual cubes were combined and weighted by their S/N. The KCWI instrument has rectangular pixels on the sky, so the 3D cubes were re-binned by interpolating in the $x$-direction to produce square pixels. The sky subtraction was then performed using a blank region within the KCWI field of view, available for all of our objects due to their extreme compactness. Finally, the wavelength range was trimmed at the red end just before the strong 5577~\AA\, sky line.

\begin{figure*}
\centering
\includegraphics[width=1.0\textwidth]{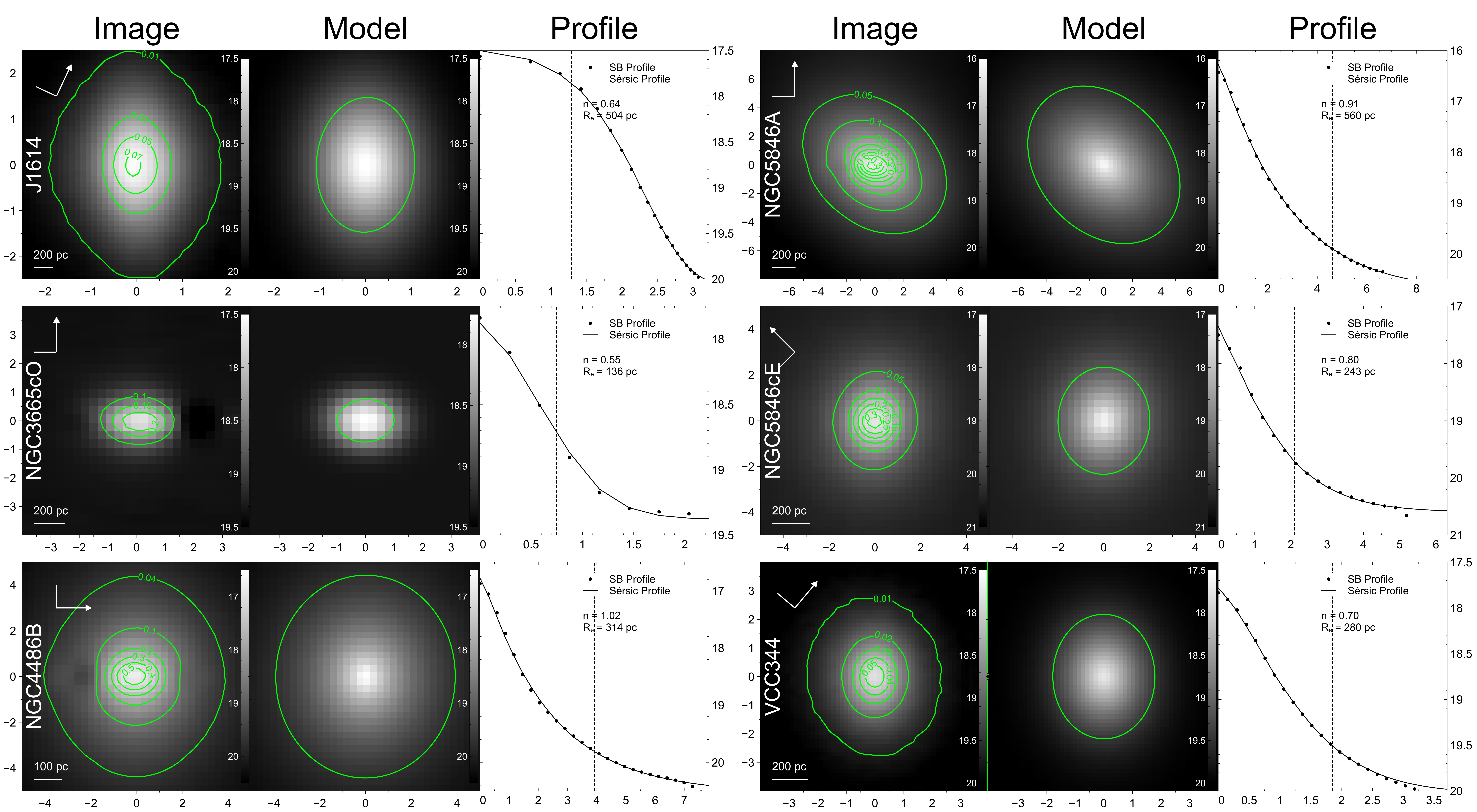}
\vspace{-0.5cm}
\caption{Image, model and 1D surface brightness profile from KCWI data for our cEs. Surface brightness is in units of mag arcsec$^{-2}$ and the axes scale in arcsec from the centre of the object. Column 1 - $V$-band surface brightness image with flux contours in units of 10$^{-16}$ erg cm$^{-2}$ s$^{-1}$ marked in green. The physical length scale is shown in the bottom left of the plot and the orientation is shown with the arrow pointing North and the short line pointing East. Note that not all the field-of-view of each configuration is shown. Column 2 - 2D model fit, with the green ellipse showing the corresponding 1$\re$ for each galaxy. Column 3 - Surface brightness and \ssc{} profile fit; the \ssc{} index and effective radius are labelled on each plot; the vertical dotted line is 1$\re$. Physical sizes are calculated assuming the distances quoted in Table \ref{tab:obs}.}
\label{fig:Sersic_maps}
\end{figure*}

\begin{figure*}
\centering
\includegraphics[width=1.0\textwidth]{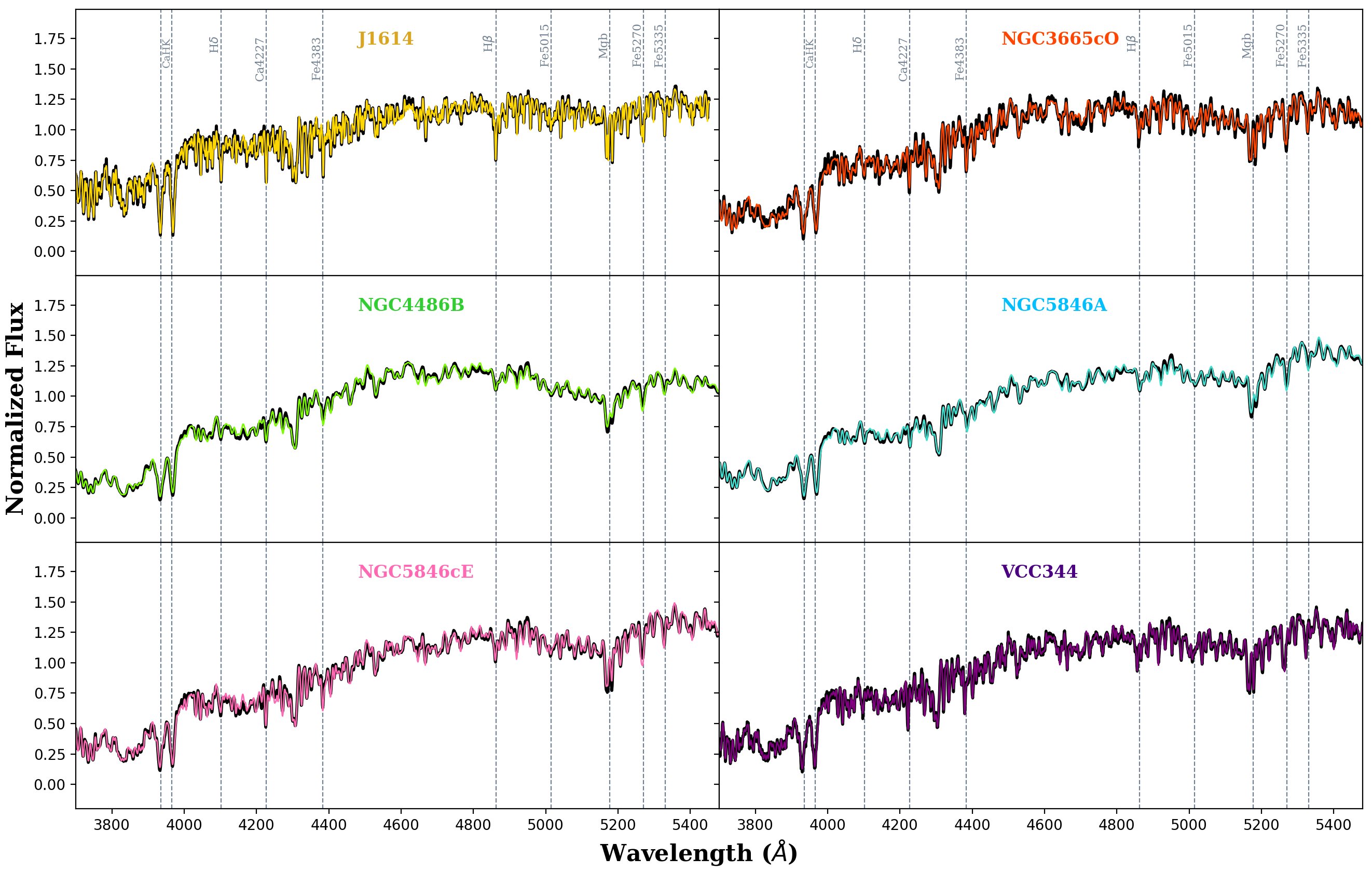}
\vspace{-0.5cm}
\caption{KCWI spectra of sample galaxies within 1$\re$ (black line). The coloured line corresponds to the \texttt{pPXF} fit for the stellar kinematics. This colour scheme will be followed throughout the rest of the paper. The main absorption line indices in the spectral range are marked and labelled with dashed vertical lines.}
\label{fig:spectra}
\end{figure*}

\begin{table*}
\caption{Summary of the measured structural properties.} 
\label{tab:params}
\begin{tabular}{lcccccccccc}
\toprule
Galaxy    &  $n$   &  $b/a$ & $\re$    & $\re$      & source           & RV        & $V_{\rm max}$  & $\sigma_{\rm e}$  & $V/\sigma_{\rm e}$ & $\lambda_{\rm e}$ \\ 
          &        &        & (")      & (pc)       &				   & (\kms)    & (\kms)         &  (\kms)           &                    &                    \\
  (1)     &  (2)   & (3)    & (4)      & (5)        & (6)              & (7)       & (8)            &  (9)              &  (10)              &   (11)             \\
\midrule
J1614      &  0.64  & 0.72 & 1.29      & 504\un36   &   this work      & 4792\un2  & 51\un2         & 36\un3            & 0.34\un0.12  & 0.35\un0.15  \\ 
NGC~3665cO &  0.55  & 0.74 & 0.74(\dag)& 136\un9    &   this work      & 1994\un6  & 6 \un2         & --                & --           & --           \\
NGC~4486B  &  1.02  & 0.89 & 2.64      & 282\un44   &   AIMSS          & 1506\un12 & 64\un2         & 170\un14          & 0.21\un0.13  & 0.19\un0.09  \\
NGC~5846A  &  0.91  & 0.76 & 4.13      & 520\un67   &   AIMSS          & 2211\un7  & 83\un3         & 155\un22          & 0.49\un0.15  & 0.47\un0.12  \\
NGC~5846cE &  0.80  & 0.86 & 2.01      & 243\un29   &   this work      & 1456\un6  & 29\un1         & 123\un9           & 0.40\un0.11  & 0.39\un0.11  \\
VCC~344    &  0.70  & 0.82 & 1.78      & 280\un39   &   this work      & 2021\un2  & 33\un2         & 85\un14           & 0.19\un0.07  & 0.22\un0.06  \\
\bottomrule
\end{tabular}
\\
\vspace{+0.2cm}
Columns: (1) Galaxy (2) \ssc{} index; (3) Minor/major axis ratio; (4) Effective radius $\re$ of \ssc{} fit, (\dag) upper limit as it does not take seeing into account ; 
(5) $\re$ (pc) with uncertainty from distance error; (6) Source for the size measurement; (7) Radial velocity (RV) from \texttt{pPXF};\\
(8) Maximum radial velocity relative to the systemic RV; (9) Velocity dispersion within 1$\re$; (10) Rotation parameter $V/\sigma$ calculated within $1 \re$; \\
(11) $\lambda$ calculated within $1 \re$.\\
\end{table*}

\section{Analysis}\label{section:ana}
\subsection{Photometry and Structural Parameters}\label{section:struct}
We generated a $V$-band image for each galaxy by averaging over a 100\,\AA{} window around 5500\,\AA{} from the flux-calibrated data cubes. These images are then converted to surface brightness (in mag\,arcsec$^{-2}$) using the constant for the Johnson $V$ filter from the SSC Magnitude/Flux Density Converter\footnote{\url{ssc.spitzer.caltech.edu/warmmission/propkit/pet/magtojy/}} and the pixel size (0.146" for the small slicer, 0.292" for the medium and large configurations). The surface brightness images were fitted with a 2D \ssc{} profile using the functionality provided in \texttt{QFitsView} \citep{Ott2012}. Figure \ref{fig:Sersic_maps} show the $V$-band images as well as the surface brightness profiles and their \ssc{} fits. Table \ref{tab:params} shows the derived \ssc{} fit parameters. To convert into physical size we use the surface-brightness fluctuation distance tabulated for the host galaxy, when available, as given in the \textit{Cosmicflows-2} database \citep{Tully2013}. For J1614, which has no clear host association, and NGC~3665 (NGC~3665cO) that has disparate redshift-independent distances, we use instead the Hubble flow with Virgo Infall + Great Attractor + Shapley corrections from the NASA Extragalactic Database\footnote{\url{http://ned.ipac.caltech.edu/}}, setting $H_{0}$ = 68 \kms per Mpc. The assumed distances used are quoted in Table \ref{tab:obs}. The uncertainty in the effective radius in parsecs is scaled from the distance uncertainty. 

We compare our results with previous literature sizes, mostly from the AIMSS compilation (from either \citealt{Norris2014} or \citealt{Janz2016}), which are based on $HST$ and SDSS imaging.
As $HST$ has better resolution, we use the literature value in those cases where a large discrepancy is found, as labeled in Table \ref{tab:params}. While we find a similar size for NGC~5846cE and J1614 (243\,pc compared to the 240\,pc and 504\,pc compared to 511, respectively), the differences for NGC~4468B and NGC~5846A are larger (314\,pc compared to 180\,pc and 560\,pc compared to 520\,pc) and therefore we use the published sizes for these two galaxies. We find that all the cEs have a low \ssc{} index ($n <$1) and a mean surface brightness at 1$\re$ of $\sim$18--18.5 (in the $V$-band), which are values similar to the cEs in \hyperlink{G+15}{G+15}.

\subsection{Stellar kinematics and stellar populations}\label{section:ssp}
We derive integrated values within 1$\re$ but also study the spatially-resolved properties applying a Voronoi tessellation \citep{Capellari2003} to our data cubes, which spatially bins the spectra to optimize the data to the required signal-to-noise (S/N) value. A S/N of 10--15~\AA$^{-1}$ has been shown to be sufficient for recovering stellar populations (see e.g.\ \citealt{CidFernandes2015}; \citealt{Citro2016}; \citealt{Constantin2019}). However, we apply higher S/N cuts when possible to provide more robust measurements. Given the different physical scale and the quality of the observations, we apply a different S/N threshold to each individual galaxy, ranging from 10 to 40. For each case we then choose the smallest S/N in order to maximize the number of bins, as long as this does not compromise the spatial information. Then the tessellated maps and the spatially-collapsed spectrum within 1$\re$ are fed into \texttt{pPXF} (Penalized Pixel Fitting; \citealt{Cappellari2004}) to obtain the stellar kinematics (radial velocity and velocity dispersion, when the instrumental resolution allows it), and the stellar populations (age, metallicity and star formation histories). We proceed with a multi-step process. We first run \texttt{pPXF} using the full MILES library of stellar templates (\citealt{Sanchez-Blazquez2006}; \citealt{Falcon-Barroso2011}), which has a nominal resolution of FWHM = 2.5~\AA\ . We initially fit the 1$\re$ spectra, while increasing the value of the additive polynomials, until reaching a value where both kinematic parameters stabilize, without using any regularization scheme. 

\begin{table*}
\centering
\caption{Summary of the measured stellar population properties}
\label{tab:pops}                      
\begin{tabular}{lcccccc}    
\toprule      
\textbf{Galaxy}  &  \textbf{Age (Gyr; SSP)} & \textbf{Age (Gyr; mass) }& \textbf{[$Z$/H] (dex; SSP)}  &  \textbf{[$Z$/H] (dex; mass)}  & \textbf{[$\alpha$/Fe] (dex)} & $M_* (10^{9}$ M$_{\odot})$\\
(1)       & (2)             & (3)        & (4)    & (5)   & (6)  & (7) \\
\midrule
J1614               &  $3.5 \pm 0.5$           &  $4.5 \pm 0.8$           & $-0.13 \pm 0.06$            & $-0.18 \pm 0.05$               &  $0.05 \pm 0.10$               &   $4.82 \pm 1.06$\\
NGC~3665cO       & $13.8 \pm 4.8$           & $11.6 \pm 0.8$           & $-0.10 \pm 0.18$            &  $0.24 \pm 0.04$               &  $0.05 \pm 0.15$               &   $0.93 \pm 0.04$\\
NGC~4486B            & $14.0 \pm 3.2$           & $12.4 \pm 0.9$           &  $0.23 \pm 0.04$            &  $0.18 \pm 0.10$               &  $0.40 \pm 0.10$               &   $5.61 \pm 1.51$\\
NGC~5846A            &  $9.8 \pm 2.1$           & $12.3 \pm 0.5$           &  $0.40 \pm 0.10$            &  $0.20 \pm 0.11$               &  $0.30 \pm 0.05$               &   $12.6 \pm 4.00$\\  
NGC~5846cE       &  $9.3 \pm 1.8$           & $12.2 \pm 0.9$           &  $0.26 \pm 0.14$            &  $0.19 \pm 0.10$               &  $0.30 \pm 0.10$               &   $2.22 \pm 0.11$\\  
VCC~344              &  $9.8 \pm 1.6$           & $10.9 \pm 0.9$           &  $0.28 \pm 0.12$            &  $0.27 \pm 0.11$               &  $0.05 \pm 0.10$               &   $8.85 \pm 0.47$\\  
\bottomrule
\end{tabular} 
\\
\vspace{+0.2cm}
Columns: (1) Galaxy; (2) Luminosity-weighted mean stellar age from absorption line indices; (3) Mass-weighted mean stellar age from \texttt{pPXF};\\
(4) Luminosity-weighted mean stellar metallicity from absorption line indices; (5) Mass-weighted mean stellar metallicity from \texttt{pPXF}; \\
(6) Mean $\alpha$-enhancement from absorption line indices; (7) Stellar mass obtained from the SFH of each galaxy.\\
\end{table*}
\begin{figure}
\centering
\includegraphics[scale=0.35]{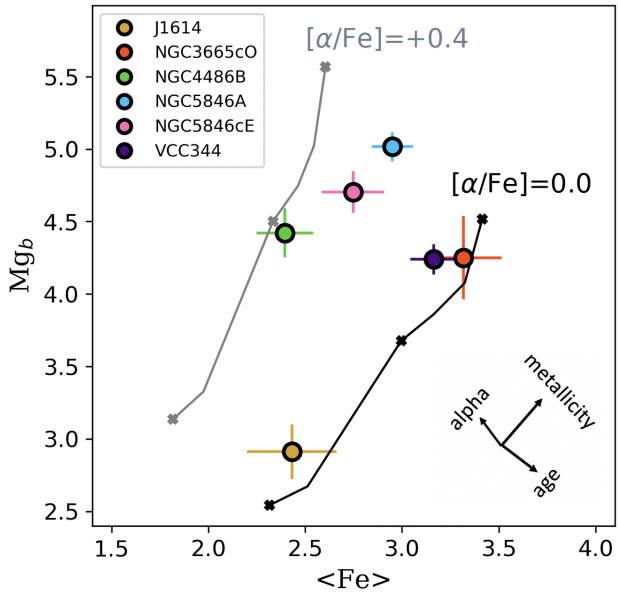}
\caption{Magnesium vs.\ iron line index plot for estimating alpha-element enhancement of cEs. SSP model index grids to obtain the $\alpha$-enhancement needed for the second step of the procedure. Two sets of the \citet{Vazdekis2016} models with different [$\alpha$/Fe] values (0.0 and $+$0.4\,dex) are shown for an age of 10\,Gyr, for simplification. Three total metallicities are marked with crosses for each SSP ([$Z$/H] $= -$2.42, $+$0.06 and $+$0.40\,dex) and the arrows shows how age, metallicity and alpha abundances vary along the model grids. The [$\alpha$/Fe] value is obtained from a basic linear interpolation between grids, rounded to 0.05\,dex. Coloured circles follow the scheme in Figure \ref{fig:spectra}.}
\label{fig:alpha_plot}
\end{figure}

This polynomial degree is then used in fitting the datacubes and deriving the stellar kinematic parameters. This step of the analysis also produces a clean, emission-free spectrum from the GANDALF routine \citep{Sarzi2017}, which is included in the \texttt{pPXF} distribution. The values obtained for the radial velocity and the velocity dispersion are shown in Table \ref{tab:params}, with the associated uncertainties obtained from running Monte Carlo simulations. From the radial velocity and the $\sigma$ we derive the spatially-integrated kinematic parameters that can provide information about the mass assembly of the galaxies. 
We derive two different parameters describing rotation dominance: the more traditional $V/\sigma$ \citep{Binney2005} and the specific stellar angular momentum ($\lambda_R$; \citealt{Emsellem2007}), computed as a function of enclosed radius along the major axis ($R$):
\begin{equation}
 (V/\sigma)^2 = \frac
 {\sum_{i=1}^{N} F_i\,V_i^2 }
 {\sum_{i=1}^{N} F_i\,\sigma_i^2};\,\,\, \lambda_{R} = \frac
 {\sum_{i=1}^{N_p} F_i\,R_i\, \textbar V_i \textbar}
 {\sum_{i=1}^{N_p} F_i\,R_i\,\sqrt{V_i^2+\sigma_i^2}}
\end{equation}
\noindent
where $F_i$, $R_i$, $V_i$ and $\sigma_i$ are the flux, radius, rotational velocity (corrected by the systemic velocity) and velocity dispersion at each spatial $i$th Voronoi bin. We use the notation ($V/\sigma$)$\rm_e$ and $\lambda\rm_e$ for the values of these properties when measured within 1~$\re$, as shown in Table \ref{tab:params}. Figure \ref{fig:spectra} shows the reduced spectra and the fit obtained in this first step for the 1$\re$ apertures, to show the quality both of the spectra obtained and of the fits.

From this point forward, we use the MILES Single-Stellar Population (SSP) library \citep{Vazdekis2010} models with the BaSTI isochrones, considering templates that range from metallicities of [$Z$/H] $= -$2.42 to $+$0.40\, dex and that cover ages from 0.03 to 14\,Gyr. Although it is known that a varying IMF will impact the derived stellar populations \citep{Ferre-Mateu2013}, the expected IMF for low-mass galaxies is Kroupa-like \citep{Kroupa2001}. We therefore use a universal Kroupa-IMF throughout this paper. This also allows for comparison with literature values. Additionally, this suite of SSP models allow for different [$\alpha$/Fe] values (either solar or $+$0.4\,dex). We therefore use the emission-cleaned spectra from the first step to measure a set of absorption-line indices such as the age sensitive H$\beta$, the set of iron indices Fe5015, Fe5270 and Fe5335 and the $\mathrm{Mg_{b}}$. These are commonly used to obtain mean luminosity-weighted ages, metallicities, [Fe/H] and more importantly, to derive [$\alpha$/Fe]. The latter is obtained from the $\mathrm{Mg_{b}}$--$\mathrm{<Fe>}$ pair, as shown in Figure \ref{fig:alpha_plot}. Here the SSP models correspond to an age of 10\,Gyr, for both the scaled-solar and the [$\alpha$/Fe]~$=+0.4$\,dex models. This age choice, made for simplification, is a valid assumption as shown in the next section. The values of [$\alpha$/Fe] for the galaxies are obtained by interpolating between the grids, rounded to 0.05\,dex. We also measure [$\alpha$/Fe] for each bin with the same approach, to see whether or not there are gradients in the abundance pattern. We do not find strong gradients and hence we adopt the [$\alpha$/Fe] value at $1 \re$ for each galaxy.

Lastly, we run \texttt{pPXF} with the set of SSP models with the [$\alpha$/Fe] value that is closest to the value from the previous absorption-line indices analysis. We thus use the models with [$\alpha$/Fe]$=+$0.0\,dex for J1614, NGC~3665cO and VCC~344, and use [$\alpha$/Fe]$=+$0.4\,dex for NGC~4486B, NGC~5846A and NGC~5846cE. The kinematics here are fixed to the values obtained in the first run. This iteration is done using multiplicative polynomials and applying a regularization value that ensures that the resulting star formation history (SFH) is the smoothest possible while maintaining a realistic fit (e.g.\ \citealt{Cappellari2010}; \citealt{McDermid2015}). To establish this regularization, we use a central bin with the highest possible S/N, and then the same regularization value is applied to the rest of the data-cube. With this last step, we obtain the mass-weighted ages, metallicities and SFHs of our cEs. We then use the SSP model mass-to-light ratio to convert the luminosity into the stellar mass for each galaxy. The uncertainties associated with the parameters in this last step are calculated as the standard deviation of the different measurements: with/without regularization, alpha-enhanced and scaled solar models. Table \ref{tab:pops} summarizes the stellar population values derived from the spectra within $1 \re$.

\begin{figure*}
\centering
\includegraphics[width=1.0\textwidth]{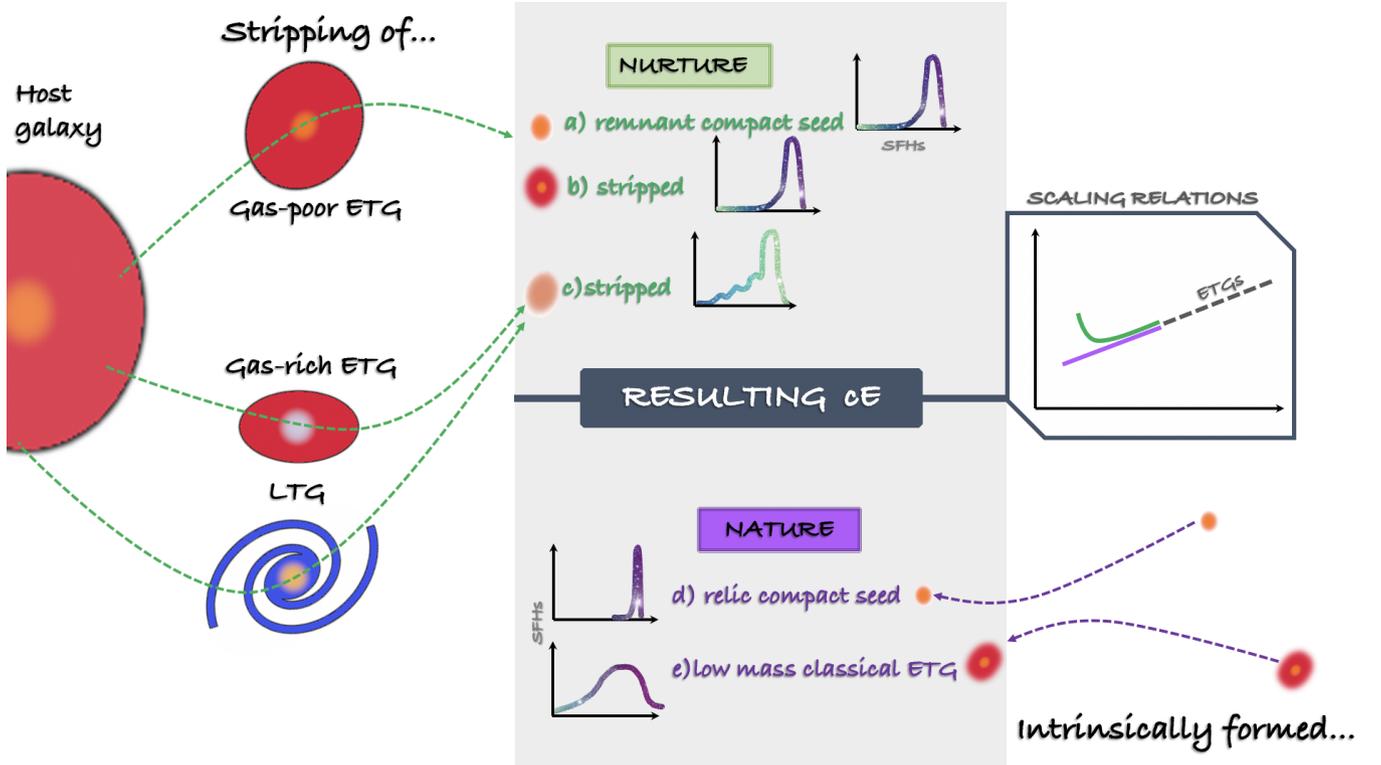}
\vspace{-0.5cm}
\caption{Schematic cartoon summarizing the different formation channels that cEs can undergo, together with their most characteristic properties such as their SFHs and scaling relations. The upper center part of the cartoon ({\it green text}) shows different outcomes of a stripping process, depending on the progenitor type (e.g.\ gas-rich spiral or ETG, or gas-poor ETG). These cEs are typically expected to deviate from most of the local scaling relations (green solid line in the right inset) and to show either bursty SFHs (time evolves to the left in the SFH diagrams) if there is gas ($c$), or very old SF episodes if they were gas-poor ETGs or the stripped remnant of the compact seed ($a$, $b$). The bottom center part of the cartoon ({\it purple text}) depicts the intrinsic scenario whereby these galaxies were formed as compact and low-mass as we see them today. These cEs should follow the scaling relations of ETGs at the low mass end (purple line in right inset), and can have either extremely old, almost SSP-like SFHs ($d$ for the relic compact seeds) or more extended ones ($e$, classical low-mass ETGs).}
\label{fig:cartoon}
\end{figure*}

\begin{figure*}
\centering
\includegraphics[width=0.83\textwidth]{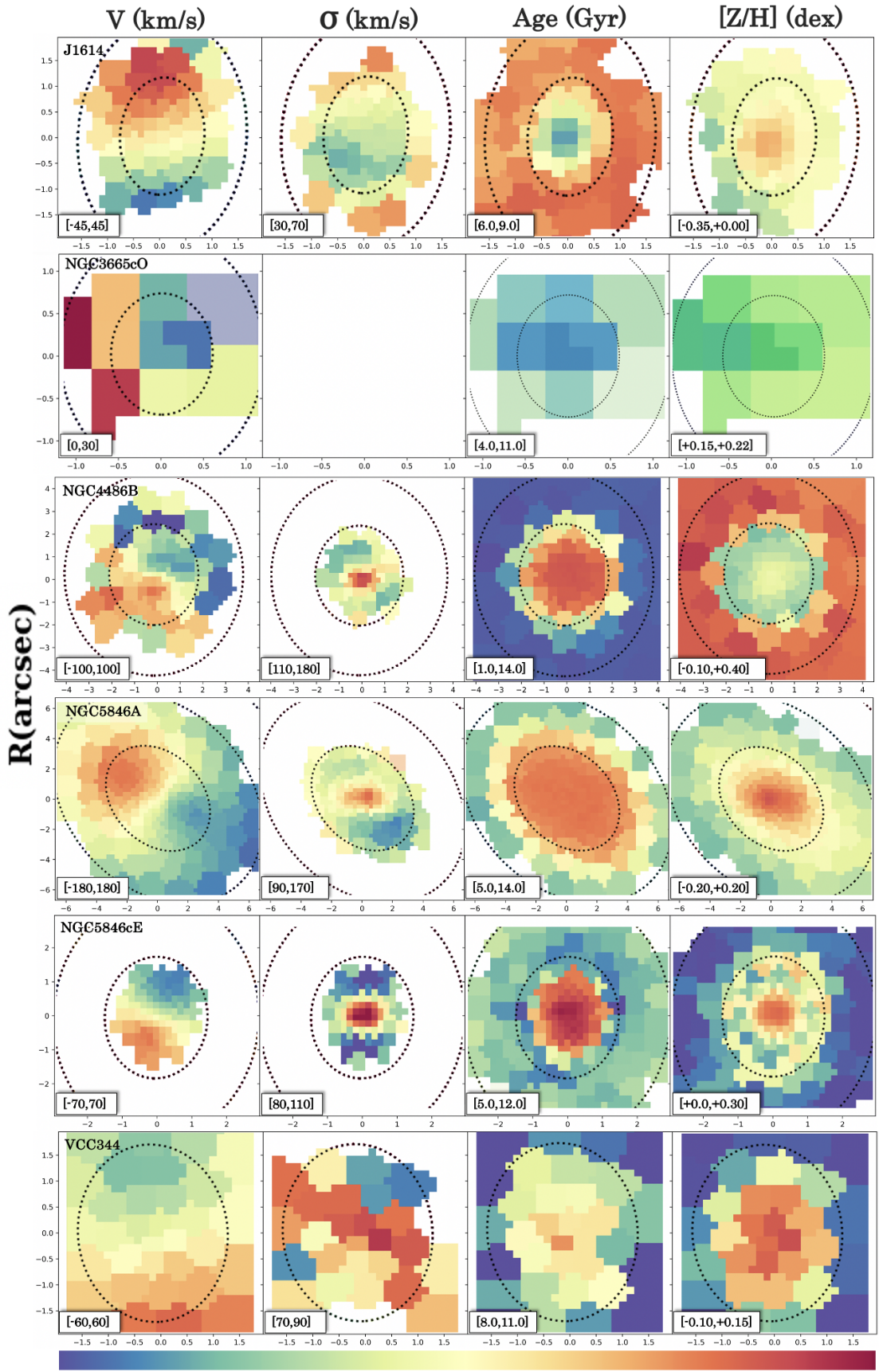}
\vspace{-0.3cm}
\caption{\textbf{2D maps of cEs from \texttt{pPXF}}: mean radial velocity (first column); velocity dispersion (second column); mean mass-weighted age (third column) and total metallicity (fourth column) for our sample of 6 cEs. Axes are in arcsec and in each panel the boxed values indicate the property range corresponding to the colour bars, with the units as in the title of each column. The dotted ellipses represent the 1~$\re$ and 2~$\re$ (where possible) isophotes  of each cE. Note that we show NGC~3665cO maps although this object is so compact that it is resolved with only a few bins.}
\label{fig:2D_maps}
\end{figure*}

\section{Results}\label{section:results}
Before we proceed to discuss the results obtained, we first summarize the main properties that can help characterize the different formation channels. We also summarize three main sets of simulations that account for the different possible formation channels of cEs. These will be used in the following sections for comparison and to draw conclusions about the formation pathways each of our cEs might have followed. 

\subsection{The formation channels of cEs}\label{section:chan}
Under the current galaxy formation paradigm, massive galaxies are thought to form in two phases. First, extreme in-situ star formation creates a superdense, compact, massive and fast-rotating galaxy at high redshift ($z>2$; e.g. \citealt{Oser2010, Oser2012}; \citealt{Schaye2015}), also known as a red nugget (e.g. \citealt{Damjanov2014}). Then, the galaxy undergoes a series of accretion events that deposit new material in the outskirts of the red-nugget, growing its size until it becomes the typical large ellipticals we see today. This means that the original red-nugget is kept hidden in the centre of present-day galaxies (e.g. \citealt{delaRosa2016}; \citealt{Ferre-Mateu2019}). However, in very rare cases, this massive compact core remains essentially untouched by avoiding the accretion phase, due to the stochastic nature of mergers (e.g.\ \citealt{Trujillo2009}; \citealt{Quilis2013}), and leading to what is referred as a massive relic galaxy (e.g. \citealt{Trujillo2014}; \citealt{Ferre-Mateu2017}). 

However, the red nugget can be separated into two photometrically distinctive parts (e.g. \citealt{Huang2013}). There is an innermost ($\sim$1kpc) compact region, representative of an early compaction phase at $z\sim$3–4 (\citealt{Hopkins2010}; \citealt{Zolotov2015}; \citealt{Martin-Navarro2019}), which contains about 15\% of the stellar mass and light of an ETG. Then there is a more disky but still compact part surrounding it ($\sim$2.5\,kpc) that comprises another additional 25\% of mass and light, and which corresponds to the quenched red-nugget formed by $z \sim 2$. Therefore, some cEs could be this seed of the earliest phases of galaxy formation, either because they remained untouched (relic compact seed), like their more massive relic counterparts (e.g. \citealt{Wellons2015, Wellons2016}), or because the fully formed massive galaxy was later stripped of both the accreted component and the most external part of the red-nugget (remnant compact seed). We emphasize here the difference between \textit{relic} and \textit{remnant}. The former makes reference to an object that is exactly as it was when formed at high redshift and that never evolved to a larger, more massive galaxy. The latter is what is left of the galaxy after its complete transformation into a massive galaxy that is later stripped, which can have suffered some alterations like gas being deposited in its center (e.g. \citealt{Ferre-Mateu2019}). In both cases, the innermost part of the galaxy ($\sim$1\,kpc; $M_{*}\sim 10^{9} \mathrm{M}_{\odot}$) is left behind as an object similar to a cE.  

Therefore, there are two main pathways, depicted in the schematic cartoon of Figure \ref{fig:cartoon}. There is the {\it nurture} path where cEs are the result of stripping processes (middle-top part of cartoon). The stripped cEs can have a variety of properties depending on what type of galaxy the progenitor was, e.g.\ gas-rich/gas-poor ETG or a LTG. Some of them could even be the remnant compact seed left behind, as described above. There is also the {\it nature} path where the cEs are formed intrinsically (middle-bottom part of cartoon) as low-mass, compact objects, this is, as the low luminsoity end of ETGs. Some could even be the untouched fossil of the early compaction phase: the relic compact seed. Some properties such as the SFHs (small inset curves) and several scaling relations (right inset of cartoon) are needed to discriminate which pathway each cE follows. Typically, cEs of a stripped origin will deviate from most of the scaling relations, as shown with the green line, and will show either very old and early SFHs if the progenitor was a massive ETG ($a,b$), or a more bursty and extended SFH if the progenitor was a gas-rich galaxy ($c$). In the alternative scenario, the cEs will follow the low-mass end of the local scaling relations (purple line) and will have either very old, almost SSP-like SFHs ($d$) or extended ones with low star formation rates, as expected for typical low-mass galaxies ($e$). In this work we consider several spatially-resolved properties in a sample of cEs with the aim of further understanding the diverse pathways depicted in this figure.

\subsection{Simulations}\label{section:simul}
\begin{itemize}
    \item \citet{Martinovic2017}: \textit{Type:} Illustris-1 cosmological hydrodynamical simulation; \textit{Assumption:} (1) stripped and (2) intrinsic, both in the vicinity of a massive galaxy in a cluster; \textit{Result:} (1) Milky Way-mass galaxies start losing their mass as they enter the cluster environment at $z \sim 2$. By $z \sim 1$ they have lost most of their mass, having sunk into the centre-most parts of the cluster, while losing their gas and dark matter, leaving a compact remnant behind as in scenario $a$. (2) they form already inside the cluster at lower redshifts ($z \sim 1$) from dense gas clouds without dark matter. They have a short formation phase and can lose some of their more loosely-bound stars, but on average are larger than the stripped cEs (scenario $e$).  
    \item \citet{Du2019}: \textit{Type:} high-resolution N-body plus gas simulations; \textit{Assumption}: formed by stripping; \textit{Result}: they start with a low-mass, gas-rich galaxy orbiting a host galaxy. The galaxy keeps some gas in the centre that produces bursty SFHs due to the combination of ram-pressure confinement and tidal stripping, creating a metal-rich compact bulge. Then the cE is quenched and suffers tidal interactions that further change its size, morphology, mass and metallicity, similar to scenario $c$, while also allowing for a wide range of alpha-abundances. They also simulate a gas-rich dwarf-like galaxy that evolves in isolation. It increases its size and stellar mass due to stellar feedback, and ends up less metal-rich than the cEs.   
    \item \citet{UrrutiaZapata2019}: \textit{Type:} numerical simulation; \textit{Assumption:} intrinsic origin by merging star clusters; \textit{Result:} merging of several star clusters with UCD-like masses, formed in the early, gas-rich Universe. All their simulations produce bound stable objects after 10 Gyr with similar sizes, surface brightnesses and velocity dispersion to some of the observed cEs, in particular the less massive and more compact ones like in scenario $d$, as they are expected to have been formed at the earliest stages of galaxy formation.
\end{itemize}

\subsection{Stellar kinematics}\label{section:kin}
The first and second column of Figure \ref{fig:2D_maps} show the rotational velocity (corrected by the systemic velocity of each cE) and the velocity dispersion measured with \texttt{pPXF}. NGC~3665cO is the only one of the 6 cEs that has a velocity dispersion well below the instrumental resolution of the configuration used, hence the missing 2D map in Figure \ref{fig:2D_maps}. Due to its small size, the smallest in our sample, and the poor seeing during its observations, we caution the reader about the other maps for this cE. While we include them in the figure to show that there are no major gradients, these will not be used to derive any gradients. Not taking into account this objects, all of our cEs show moderate rotations with $V_{\rm max}$ of around 20--60\,\kms. The notable exception is NGC~5846A, which shows a larger rotation value of $\sim$80\,\kms. The second column of Figure \ref{fig:2D_maps} presents the 2D velocity dispersion maps. It shows that NGC~4468B, NGC~5846A and NGC~5846cE have peaked central velocity dispersions, whereas both VCC~344 and J1614 have less spatial structure, with mostly flat distributions. 

\begin{figure}
\centering
\includegraphics[scale=0.6]{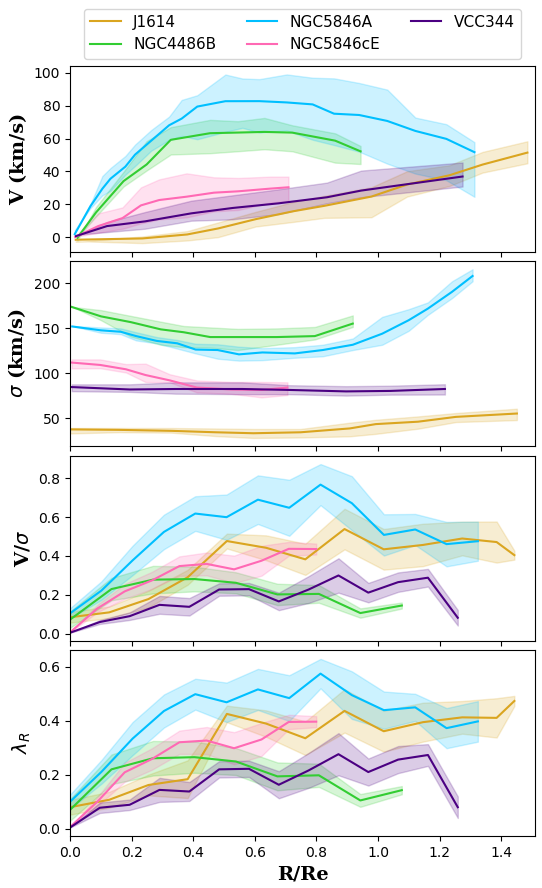}
\vspace{-0.3cm}
\caption{Azimuthally averaged radial kinematic profiles of our cEs. Showing the radial distance from the centre, the first kinematic orders (RV corrected by the systemic velocity in the top panel, and the stellar velocity dispersion, $\sigma$, in the second panel) are shown together with the V/$\sigma$ and $\mathrm{\lambda_R}$ values. The colour scheme follows Figure \ref{fig:spectra}, with the crosses marking the mean value at a given radius and the filled area shows the uncertainty of the measured property, computed as the 1$\sigma$ value of the mean. Note that the values for NGC~3665cO are not shown, since the $\sigma$ for this galaxy is lower than the resolution of the configuration used. }
\vspace{-0.3cm}
\label{fig:grads_kinem}
\end{figure}

\begin{figure*}
\centering
\includegraphics[scale=0.55]{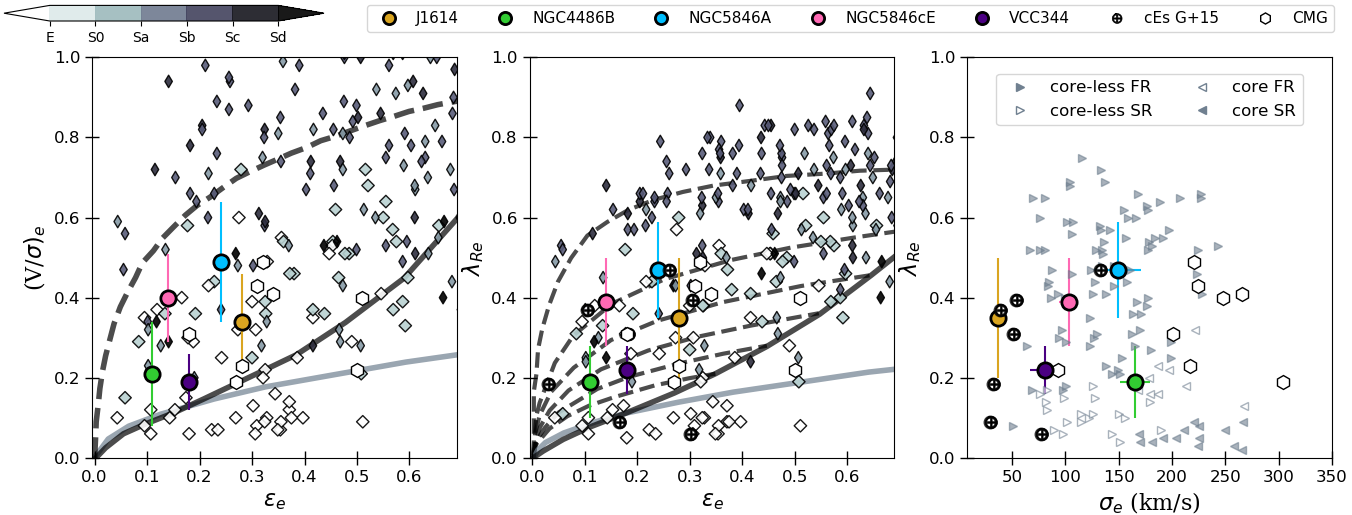}
\vspace{-0.3cm}
\caption{\textit{Left:} $V/\sigma$--$\epsilon$ relation within one effective radius to determine whether our sample of cEs (coloured dots as in Figure \ref{fig:spectra}) are fast or slow rotators. For comparison, a sample from the CALIFA survey is shown (diamonds, colour-coded and shaped by different morphological types; \citealt{Falcon-Barroso2019}) and CMGs from \citet{Ferre-Mateu2012} and \citet{Yildirim2017}. The solid grey curve shows the empirical division that separates fast (above) and slow (below) rotators \citep{Emsellem2007}. The solid black curve describes the relation $\delta = 0.7\,\epsilon_{\rm intr}$ from \citet{Cappellari2007}, while the black dashed line shows this relation for a maximum inclination. \textit{Middle:} Similar to the left panel but showing the $\lambda_R$--$\epsilon$ relation from \citet{Emsellem2007} as the solid grey line. The solid black curve describes the edge-on relation for fast rotators from \citet{Cappellari2007}) whereas the dashed curves show how this relation varies with different inclinations. In this panel we also include the sample of Virgo cEs by G+15 (crossed circles). All of our galaxies are above the curve that separates slow from fast rotators. \textit{Right:} $\lambda_{R}$--$\sigma$ relation for our cEs and for a sample of giant ETGs \citep{Krajnovic2020}, which shows that our cEs are more compatible with being core-less fast rotators.}
\label{fig:rotatros}
\end{figure*}

To better quantify the different trends in the galaxies, we derive azimuthally averaged radial profiles of the kinematic values, as shown in Figure \ref{fig:grads_kinem}. Three galaxies have slowly rising rotation profiles reaching low rotational velocities of $\sim$25--45\,\kms, while the other two galaxies (NGC~5846A and NGC~4486B) rise quickly with a peak rotation at $\sim$0.5--0.7$\re$ of around $\sim$90 and 60\,\kms, respectively, which then decline at larger radii. In terms of their velocity dispersion, NGC~4486B, NGC~5846A and NGC~5846cE have a high value in their centers that then decreases by around $\sim$30\,\kms\ out to $1 \re$, although both NGC~5846A and NGC~4486B show signs of a rising profile afterwards. VCC~344 shows a virtually constant velocity dispersion of $\sim$85\,\kms\ out to more than 1$\re$. J1614, the cE with the lowest measured velocity dispersion, also remains virtually flat up to $\sim 0.8 \re$, where a slight increase is seen. Such a $\sigma$ profile was obtained for the simulated cEs of \citet{UrrutiaZapata2019}, under a intrinsic origin. The two bottom panels of Figure \ref{fig:grads_kinem} show the ($V/\sigma$) and $\lambda_R$ trends, obtained as described in Section \ref{section:ana}. In both panels, NGC4486B and VCC~344 show a gentle, almost constant, increase up to 1~$\re$ and beyond. J1416 shows a steeper gradient although it plateaus around $\sim$0.5\,$\re$. NGC5846cE shows a similar trend to J1416 but we do not reach as far for this object and thus it is unclear whether it would plateau or keep rising. NGC5846A shows the most steep profile, peaking at 0.5~$\re$ with relatively high rotation values (V/$\sigma\sim$0.6 and $\lambda_{\rm R}\sim$0.5). 

Comparing to the kinematic profiles for the simulated cEs of \citet{Du2019}, they found $V/\sigma$ profiles rising gradually to $\sim$~0.3--0.5 values at the equivalent to $\sim 2 \re$, and then flattening off at larger distances. This means that if we consider the same range of our cEs ($<$1.5$\re$), our profiles are compatible with the trend of reaching a plateau that they find. Note, however, that if our observations were de-projected, the rotation could be higher than in the simulations. We can also compare our observed cE kinematic tracks with different types of galaxies ranging from E/S0s to late-type spirals, both from observations and simulations (e.g. \citealt{Naab2014}; \citealt{Falcon-Barroso2019}). However, due to the different physical scales between large massive galaxies and the small compact galaxies considered in this work, we caution that a direct comparison of these tracks is not straightforward. One should be comparing only the inner $\sim$0.2\,$\re$ parts of massive galaxies, where basically no differences within galaxy types are yet seen. Looking at the simulated galaxies from \citet{Naab2014} the only clear result is that cEs follow the same trends as fast rotators. But if we compare the tracks with the those of massive compact galaxies (CMGs; e.g. \citealt{Ferre-Mateu2012}; \citealt{Yildirim2017}), we find that cEs are very similar to those in \citet{Yildirim2017} below 1\,$\re$. This is the first hint that compact galaxies at different mass ranges might be connected.

To further analyze the kinematics of our cEs, we show in Figure \ref{fig:rotatros} (left and middle panels) the overall values of $V/\sigma$ and $\lambda_R$ within $1 \re$, related to the ellipticity of the galaxy, $\epsilon$. The figure shows the values for our cEs and those in Virgo from \hyperlink{G+15}{G+15}, when available, although the latter are measured within $\re$/2. It is in any case apparent that all of our cEs, as well as most of the cEs from \hyperlink{G+15}{G+15}, are fast-rotator galaxies, having relatively high rotation for their low ellipticities (above the grey curve in the middle panel). Three of them have values around $\lambda_{\rm R}\sim$0.4, while NGC~4486B and VCC~344 have lower values around 0.2. We can now attempt to determine the type of progenitor these cEs could have had by comparing the loci of different types of galaxies \citep{Falcon-Barroso2019}. These are colour-coded by their morphological type and have different symbols: E/S0s are shown with light wide diamonds while the spiral types are plotted as darker narrow diamonds. The massive compact counterparts from \citet{Ferre-Mateu2012} and \citet{Yildirim2017} are shown as white pentagons. At first glance, the majority of galaxies in the region of our cEs are fast rotating ETGs (both extended or compact), although some Sa-Sb galaxies are located near the cEs with higher rotation values (e.g. NGC~5846A, NGC~5846cE and J1614). We caution again with the size comparison issue described above. If we use the values at 0.5\,$\re$ from \citet{Falcon-Barroso2019}, the control sample would shift towards lower rotation values. The results will not change for the low $\lambda_{\rm R}$ cEs, but they will be even more similar to the spiral types. We therefore need additional information, such as the one from the stellar populations to give better constrains on the progenitor type.
The right panel of Figure \ref{fig:rotatros} shows the location in the ($\lambda_{\rm e}$, $\sigma$) plane \citep{Krajnovic2020} within 1$\re$, that separated galaxies between core or core-less objects. Our results suggest that cEs are more compatible with being coreless fast rotators, with the exception of three cEs from \hyperlink{G+15}{G+15} (which include their two slow rotators).

\begin{figure*}
\centering
\includegraphics[width=0.95\textwidth]{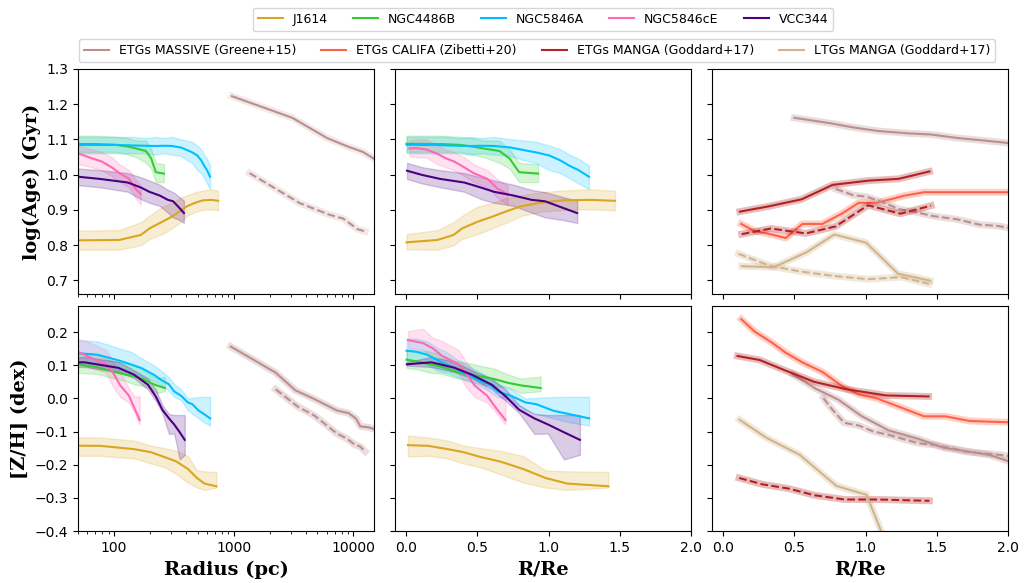}
\caption{Age and metallicity radial profiles of cEs. The mean mass-weighted age and total metallicity obtained from \texttt{pPXF} are shown as a function of radius both on a physical scale (left column) and as $R/\re$ (middle column). Colours are the same as in Figure \ref{fig:spectra}. The last column shows the profiles versus $R/\re$ for giant ETGs and LTGs in the literature: CALIFA \citep{Zibetti2020}, MASSIVE \citep{Greene2015} and MaNGA \citep{Goddard2017}. The solid line represents their most massive bin, whereas the dashed line corresponds to the least massive one in terms of velocity dispersion. Most of the cEs show a declining age gradient, although J1614 shows an increasing trend, with the centre being much younger. The metallicity profiles are mostly declining, with the majority of cEs presenting high metallicities in their centers, except for J1614.}
\label{fig:grads_pops}
\end{figure*}

\subsection{Stellar Populations}\label{section:ssp2}
Figure \ref{fig:2D_maps} also presents the mass-weighted age and metallicity maps derived from our spectral fitting procedure (third and fourth columns). Only J1614 is younger in the centre, while the rest show very old stellar ages ($>$10\,Gyr) in their centers that decrease or remain virtually constant with radius (e.g.\ NGC~3665cO). Table \ref{tab:pops} summarizes the mean values derived from both the line-index approach (luminosity-weighted SSP-equivalent values) and the full spectral fitting (mass-weighted average ones) for the 1$\re$ spectra.  

Figure \ref{fig:grads_pops} shows the azimuthally averaged age (top panel) and total metallicity (bottom) trends for the cEs, both as a function of physical radius (left) and relative to $\re$ (middle panel). In all but one case, the age shows a decline outward of about 2--4\,Gyr, while the metallicity decreases more steeply. NGC~5846A and NGC~4486B are old ($\sim$13\,Gyr) throughout their structure, although after 1$\re$ both galaxies become $\sim$1\,Gyr younger. NGC~5846cE shows a steady decrease of its age from $\sim$12\,Gyr in the centre to $\sim$8\,Gyr by almost 1$\re$. VCC~344 has a similar profile although it shows slightly younger central ages ($\sim$10\,Gyr). The metallicity profiles of these four galaxies are very similar, starting with super-solar metallicities in the centers and decreasing to sub-solar values of around $-$0.1\,dex outward of 1$\re$. Here it is even more clear how different J1614 is to the rest of cEs: it has a rising age profile (from $\sim$6\,Gyr in the centre to 8\,Gyr by $1.5\,\re$) and shows low metallicities throughout its entire structure ([$Z$/H]$\sim -0.15$\,dex). 

This figure also includes age and metallicity profiles from different surveys, shown in the right panel. We caution that as we have seen in the previous sections, these galaxies should be compared to the innermost regions of massive galaxies or compact massive ones. Unfortunately only the MASSIVE survey \citep{Greene2015} provides values on a physical radius scale that can be directly compared to our galaxies (left panel). For the other surveys, CALIFA \citep{Zibetti2020} and MaNGA \citep{Goddard2017}, we show them in a $R/\re$ scale (right panel). Therefore we here compare the trends of our cEs qualitatively rather than the absolute values. For the MaNGA and MASSIVE galaxies, the most massive of their sample (continuous line) and the least massive one (dashed line) are plotted separately, when available. All cEs have a similar shallow trend towards younger ages at large radius, with the exception of J1614. This object follows the slightly rising profiles (older at larger radius) of the ETGs from CALIFA and MaNGA (in particular the low-mass MaNGA objects). This is also seen for the metallicity, where J1614 is more metal-poor than the other cEs. The rest of cEs follow similar trends to the ETGs in all the three surveys, although the CALIFA metallicities are typically higher. As with the ages, none of the cEs seem to have profiles similar to the literature late-type galaxies (LTGs), which typically show low metallicities and young ages in their centers ($\sim\,-$0.1\,dex and $\sim$5\,Gyr). This means that the cEs would require significant chemical and temporal evolution to be compatible with a progenitor of this type. We also compare our stellar population trends with the least massive of the compact ellipticals of \citet{Yildirim2017}, as one of them is actually a cE ( \hyperlink{FM+18}{FM+18}). We find again that the trends of our cEs are compatible with the central parts of the massive counterparts. 

To further explore the possible origins and progenitor types, we now look at the spatially-integrated SFHs of these galaxies. There are three types of SFHs seen in cEs (\hyperlink{FM+18}{FM+18}), each one more indicative of a different origin (also summarized in Figure \ref{fig:cartoon}). First, stripped galaxies whose progenitor was a massive ETG will have mostly uniformly old ages, with the stars being formed early on as the central part of a red-nugget (e.g.\ \citealt{Oser2010}; \citealt{Damjanov2014}). They tend to have early and peaked SFHs similar to those seen in these red-nuggets (e.g.\ \citealt{Ferre-Mateu2017}; \citealt{Martin-Navarro2019}). If there was any gas left during the interaction, some of the younger stars formed could arrive into the core of the galaxy (e.g.\ \citealt{Ferre-Mateu2019}). Second, stripped galaxies whose progenitors were spiral-like or gas-rich E/S0s, with large amounts of gas available, will show bursty and extended SFHs. Third, the galaxies that are not the result of stripping, but are intrinsically low-mass, compact galaxies, will either have very extended SFHs with low levels of star formation through them \citep{Thomas2005} or almost single-burst ones if they are the relics of the compact seed. Therefore, the SFHs of these galaxies can provide us with definitive clues about their origins and plausible progenitors. 

\begin{figure*}
\centering
\includegraphics[width=1.0\textwidth]{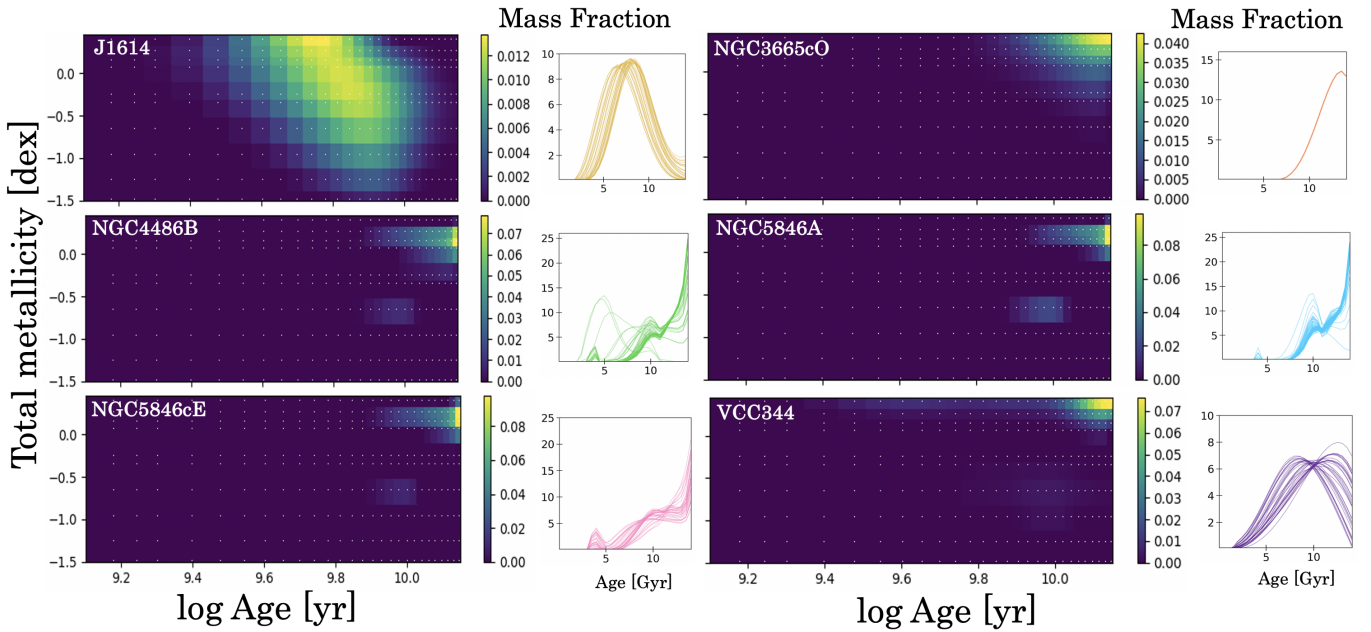}
\vspace{-0.5cm}
\caption{The age--metallicity relation and the fraction of stellar mass created (normalized to 1) at a given time as the colour-scale for each cE, as labeled. The small panel to the right side of each age--metallicity panel corresponds to the SFHs derived for all the bins with good S/N from the galaxy, with the mass fraction now normalized to 100. We remove the ages and metallicities where there is no contribution in any of the galaxies. These SFHs follow the regularisation scheme described in Section \ref{section:ssp}. It is clear that J1614 has a strikingly different SFH than the rest of the cEs. NGC~4486B, NGC~5846A and NGC~5846cE have very similar SFHs and therefore, most likely have the same type of progenitor galaxy, while NGC~3665cO and VCC~344 also have a similar SFH.}
\label{fig:sfhs_plot}
\end{figure*}

Figure \ref{fig:sfhs_plot} shows the age--metallicity relation for these galaxies, coloured by the fraction of stellar mass created at a given time. It also shows for each cE (except for NGC~3665cO) the SFHs in each bin, in order to see if there are distinctive stellar populations throughout the galaxy. It is straightforward to see that J1614 is, again, strikingly different to the other cEs. While the majority of cEs are mostly old and metal rich, J1614 is younger and less metallic throughout its entire structure. This different SFH, coupled with the information from the other properties, suggests that J1614 is an intrinsically low-mass cE. NGC~3665cO and VCC~344 both show SFHs that are similar to what is seen for ETGs of \mstar$\sim 10^{10-10.5}$\msun\, (e.g.\ \citealt{Thomas2005}; \citealt{McDermid2015}). The other cEs (NGC~4486B, NGC~5846A and NGC~5846cE) show instead very early star formations but also secondary bursts that extend down to 2--4\,Gyr ago. This is suggestive of progenitors that were either spiral-like or gas-rich ETGs, representing star forming episodes ignited during the stripping process. Considering the age and metallicity trends discussed above, we suggest that the latter must be the case. This is further reinforced by the SFHs of the simulated cEs from \citet{Martinovic2017}. The authors showed that their cEs formed via a stripping process show very early and peaked star formation episodes, while those formed intrinsically show more sustained star forming rates with rather extended SFHs.

\subsection{Implications for the cE formation channels}\label{section:disc}
\begin{figure*}
\centering
\includegraphics[width=1.\textwidth]{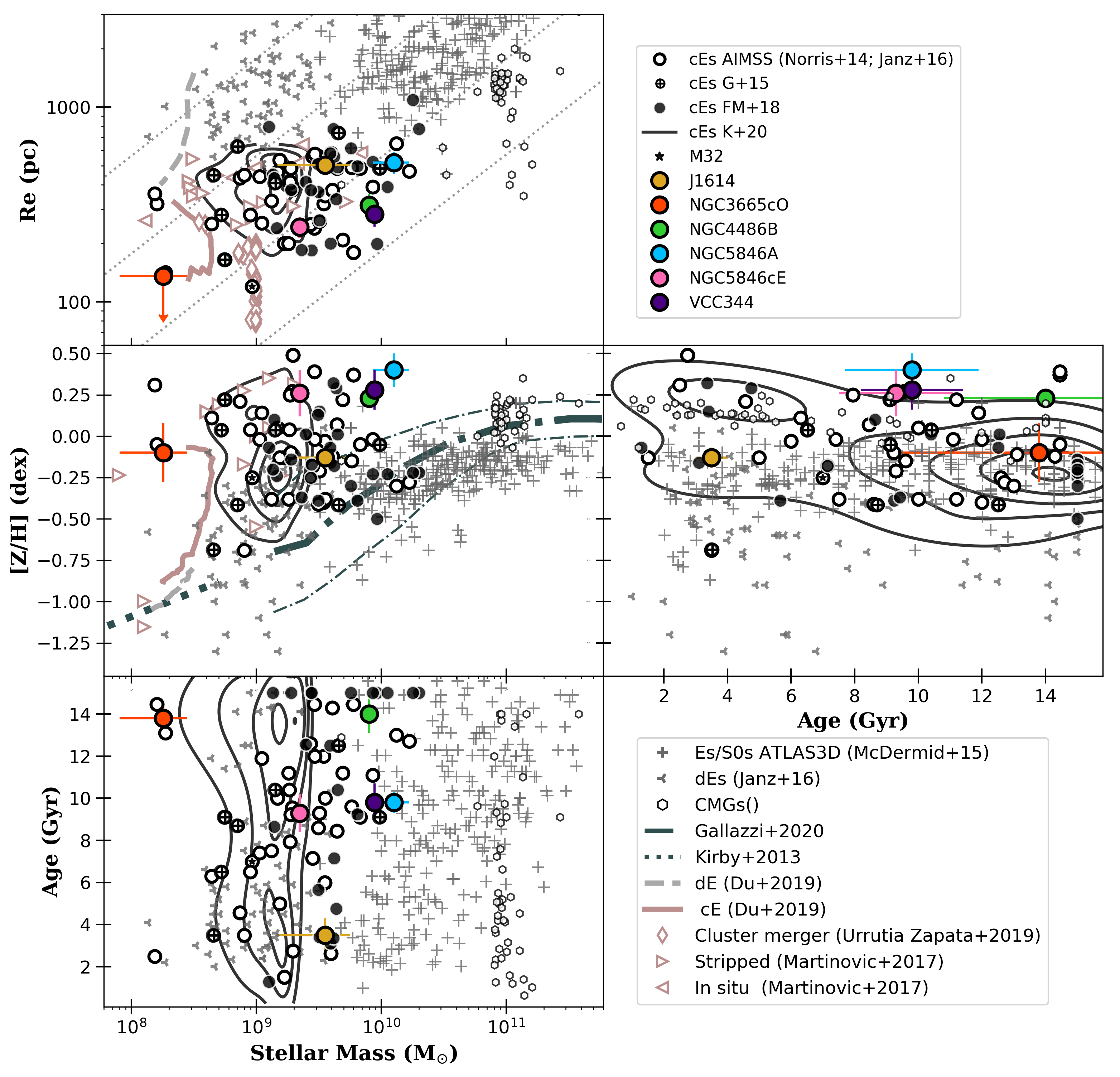}
\vspace{-0.3cm}
\caption{Scaling relations for cEs. Different types of galaxies are shown for comparison, as labeled in the legend; E/S0s (crosses), dEs (squares) from \citet{Janz2016} and \citealt{McDermid2015}, and CMGs from several compilations (see text); cEs of different samples with circles: from the AIMSS compilation of \citet{Janz2016} (white filled, with M32 as a black star), from FM+18 (black filled), from G+15 (crossed circles), and our sample (coloured dots as in Figure \ref{fig:spectra}). The cEs from the SDSS sample of K+20 (private communication) are shown as density contours We are using the line-index values to allow for comparison with the literature values and consistency. \textit{Top}: The well known mass--size relation, showing the cE regime as a continuation of the more massive ETGs. The results of different simulations are also shown: the low-mass, intrinsic cEs from \citet{UrrutiaZapata2019}, which form as the result of star cluster mergers (salmon open diamonds), the cEs from \citet{Martinovic2017} (either formed like intrinsic objects or the result of stripping, salmon open triangles) and the stripping time tracks of \citet{Du2019} for both cEs (solid salmon) and dEs (dashed grey). We also show lines of constant surface stellar mass densities from $\Sigma_{\rm eff}\sim$ 10$^{8}$\msun\,kpc$^{-2}$ to  10$^{11}$\msun\,kpc$^{-2}$ (dotted lines, increasing from left to right). \textit{Middle-left}: Mass--metallicity relation for local massive galaxies \citep{Gallazzi2020} (dashed black lines) and low mass ones \citep{Kirby2013} (dotted black line). The stellar populations used in this figure are those from the line index method, to compare with the literature.\textit{Middle-right:} Age--metallicity relation, showing a hint for older cEs being more metal-poor than younger cEs. \textit{Bottom}: Mass--age relation, showing no clear correlation for these parameters but possibly two populations of cEs: a very young and a very old one.}
\label{fig:relations_plot}
\end{figure*}

Figure \ref{fig:relations_plot} shows the main relations for the studied properties. In each panel, different types of galaxies are shown: E/S0s and dEs from \citet{Janz2016} (partially curated from ATLAS3D; \citealt{McDermid2015}), cEs from AIMSS (from both \citealt{Norris2014} and \citealt{Janz2016}), cEs from the Virgo cluster (crossed circles; \hyperlink{G+15}{G+15}), cEs from \hyperlink{FM+18}{FM+18} (black circles), the SDSS DR12 compilations of \hyperlink{K+20}{K+20} (density contours; private communication), and this current work cEs (coloured circles). When available, the properties of simulated cEs are also shown from \citet{Martinovic2017}, \citet{Du2019} and \citet{UrrutiaZapata2019}. We also include the location of CMGs (\citet{Trujillo2009}; \citet{Yildirim2017} and \citet{Spiniello2020}).

The top panel presents the well-known mass--size relation. It shows the split after the E/S0s into the realm of the diffuse dwarf objects (dEs), with typical sizes of $\re \sim$900--3000\,pc, and the realm of the compact galaxies (cEs), with sizes typically below 900\,pc. We remind the reader that these limits can vary from work to work, which can introduce some bias in the sample. For example, in \hyperlink{K+20}{K+20} the objects were selected to be smaller than 600\,pc, which means that that they are missing some of the larger (and thus more massive) cEs. This was done to avoid those objects that are in the borders separating each galaxy type. Nonetheless, we note that all but one of the cEs considered here have sizes below 800 pc. The larger cE was already reported in \hyperlink{FM+18}{FM+18} to be a low-mass ETG in the process of becoming a cE. In terms of mass, we see that the bulk of observed cEs tend to be more massive than the simulated ones. However, they are all less massive than the characteristic mass scale of 3$\times$10$^{10} $M$_{\odot}$ that marks the transition to the realm of elliptical galaxies (e.g. \citealt{Cappellari2016}). In this case, this is due to an observational bias. Obtaining high quality spectra for these objects is hard and time consuming due to their low luminosities and they will most likely not have SDSS spectra available (Ferr\'e-Mateu et al., in prep). Therefore it is possible that we are only observing the brighter, and thus more massive, tail of this family of objects (\mstar $\sim 10^{9-10}$\msun). Objects like NGC~3665cO, with \mstar $\sim 10^{8}$\msun, are thus rarer to find. Nonetheless, they are crucial to better understand the border regions at the low-mass end. To better address these selection biases, this panel also shows lines of constant surface stellar mass densities $\Sigma_{\rm eff}=$\mstar($< \re)/\pi\,\re^2$, ranging from $\sim$ 10$^{8}$\msun\,kpc$^{-2}$ to the maximum surface of $\sim$ 10$^{11}$\msun\,kpc$^{-2}$ (increasing to the right; \citealt{Hopkins2010}). All the cEs in our sample follow the line of $\Sigma_{\rm eff}\sim 10^{10}$\msun, similarly to the Es/S0s at the high mass end and similar to the well-known cE, M32 (marked with a star inside the white circle). Instead, the bulk of cEs from \hyperlink{G+15}{G+15} and \hyperlink{K+20}{K+20} follow lines of lower surface mass density. This confirms that we are, indeed, studying the most compact regime of cEs.  

From the simulations, \citet{Martinovic2017} find that on average those cEs that are the result of stripping are more compact than those that formed intrinsically, shown by the two different open triangle symbols. This is in agreement with what we find for our sample, with galaxies being mostly compact with the exception of NGC~5846cE and J1614. The latter is larger, possibly due to its intrinsic origin that we are suggesting. The cEs simulated by \citet{UrrutiaZapata2019} (open diamonds) are on average less massive and more compact than the bulk of observed cEs but this is the result of their prescription. They aim to obtain cEs in a similar way to the formation of the most massive UCDs: by the merging of stellar clusters. Therefore, their objects will cover only the low-mass end of cEs. As mentioned above, these border regions are very interesting, and are the loci of observed cEs like NGC~3665cO. We also show the stripping time tracks by \citet{Du2019}. Both their simulated dEs (grey dashed line) and cEs (solid salmon line) start with a similar mass and size (2$\times$10$^{8}$M$_{\odot}$ and $\sim$350\,pc) but then with time they follow different paths. The isolated dwarf increases the mass and size to end like a typical dE, while the cE first builds up its mass via a bursty SFH leading the compaction phase. Once quenched, it decreases in size and mass by losing the more external, loosely-bound stars. It ends up like a cE of a similar mass and size to NGC~3665cO, less massive than the bulk of (all) observed cEs. 

\begin{figure*}
\centering
\includegraphics[width=1.0\textwidth]{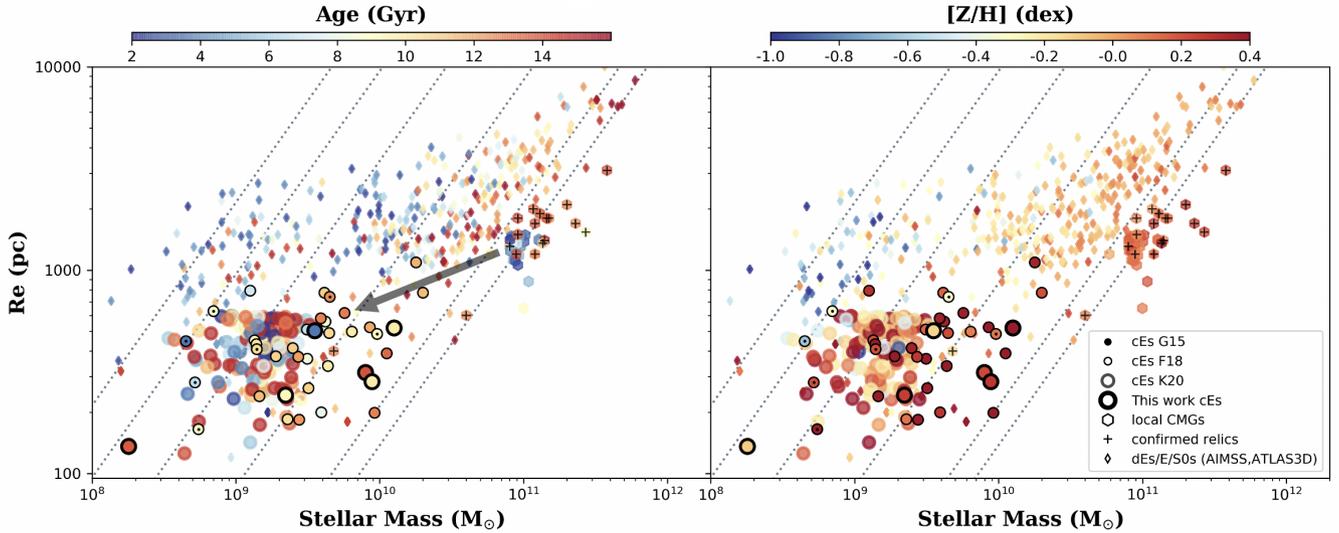}
\vspace{-0.2cm}
\caption{Galaxy size vs.\ stellar mass, similar to the top panel of Figure \ref{fig:relations_plot}, but now with all objects colour-coded by their stellar populations properties: stellar age (left panel) and total metallicity [$Z$/H] (right panel). The control sample of dEs/E/S0s are now plotted as diamonds, and the pentagons correspond to a sample of compact massive galaxies (crossed if they are confirmed massive relic galaxies; \citealt{Trujillo2009}; \citealt{Ferre-Mateu2012,Ferre-Mateu2015,Ferre-Mateu2017}; \citealt{Yildirim2017}; \citealt{Spiniello2020}). Lines of iso-velocity dispersion obtained directly from the Virial theorem are shown, ranging from 20 to 250 \kms.
The cEs in general have similar ages to the more massive ETGs, but their metallicities do not follow the lines of iso-velocity as suggested by G+15. They tend to have higher metallicities that are similar to those seen in their massive compact counterparts.}
\label{fig:relations_pops}
\end{figure*}

The left panel in the second row of Figure \ref{fig:relations_plot} shows another characteristic relation, the mass--metallicity one. The local relation for massive ETGs of \citet{Gallazzi2020} is shown in a dashed line, whereas the one for low-mass objects of \citet{Kirby2013} is a dotted line. The time tracks from \citet{Du2019} are also shown, for both stripped cEs and dEs. Their simulated dEs end close to the low-mass relation of \citet{Kirby2013}, whereas their stripped cEs follow an upward track that leaves them as outliers of the relation. This is consistent with objects that have been stripped of their stars: they lose the stellar mass but retain the original central metallicity, thus left as outliers in this relation. The main result in \hyperlink{K+20}{K+20} was that galaxies with typically higher metallicities corresponded to those associated to a host galaxy (hence with higher chances of being stripped objects), while those in the field had typically lower metallicities that are close to the low-mass end relation for ETGs, suggesting that they could have been formed intrinsically. This suggests that local environment could be a crude proxy to differentiate between stripped/intrinsic origins. However, both \hyperlink{FM+18}{FM+18} and \hyperlink{K+20}{K+20} suggested that additional properties are necessary to truly disentangle both origins.  For example, some of the field galaxies had been found to have unexpectedly high metallicities, which would indicate stripping, but that seems unfeasible without a nearby host. On the contrary, some cEs in clusters present very low metallicities, suggesting that they could be of the intrinsic origin. Looking at this panel it is straightforward to see that the bulk of observed cEs are typically outliers in this relation, and thus prone to be stripped objects. This is also the case for the majority of the simulated cEs of \citet{Martinovic2017} via stripping. Note, however, that a handful of cEs are closer to the local relation of massive ETGs, suggesting an intrinsic origin instead. The cE from our sample that is closest to the relation is J1614, which is the only cE without a host galaxy associated to. This reinforces our suggestion that this could be an intrinsic object.

Nonetheless, processes like mergers with gas-rich galaxies or strangulation can also produce high metallicities. In order to better understand the galaxy's origin, we need further information from the stellar ages. Young ages can indicate both an extended SFH but also recent star formation, while old ages will be indicative of gas-poor galaxies that formed early on. The bottom panel of Figure \ref{fig:relations_plot} shows the age--stellar mass relation. There is not a clear correlation between these two parameters, as cEs at all stellar masses cover the entire range of stellar ages. From the \hyperlink{K+20}{K+20} density contours there is a hint of a dichotomy in the age distribution, with a small group of young cEs and a larger one with old ages. This is further seen in the middle right panel of the figure, for the relation between the age and the metallicity. Following the density contours and the bulk of published cEs, these seem to be typically old and slightly more metal-poor than the small fraction of young cEs, which tend to be slightly more metallic. In fact, cEs have metallicities that are more similar to E/S0s (from $\sim-$0.2 to $+$0.2\,dex) while dEs are typically more metal-poor ([$Z$/H]$\sim -$0.7\,dex). Furthermore, the bulk of young cEs have the same ages and high metallicities as the young CMGs in \citet{Trujillo2009} and \citet{Ferre-Mateu2012}, indicating that these could be the result of stripping one of these more massive counterparts. In this case, the high metallicities could be associated to the recent episodes of star formation seen in \citet{Ferre-Mateu2012}. Note that only a few cEs cover the region of young ages and low metallicities, with J1614 in the middle of it. For these young cEs a more extended SFH is expected, similar to the ones seen for low-mass ETGs (e.g. \citep{McDermid2015}), which is in agreement with the SFH obtained in Figure \ref{fig:sfhs_plot}. On the other side of the relation, we have galaxies like NGC~3665cO, with a low metallicity similar to J1416 but extremely old ages. In this case, the object could be the result of stripping a larger, gas-poor galaxy, similar to what might have happened for NGC~4486B. For the other three cEs in our sample (VCC~344, NGC~5846cE and NGC~5846A), with very high metallicities, relatively old ages ($\sim$10\,Gyr) and bursty SFHs (Figure \ref{fig:sfhs_plot}), we suggest that their progenitor galaxy was more likely a gas-rich galaxy.  

It is clear at this point that regardless of the caveats mentioned before, a large number of cEs have a clear nurture origin. Whether or not this is the main mechanism to form cEs at all stellar masses is still to be decided. Following \hyperlink{G+15}{G+15} and \hyperlink{FM+18}{FM+18}, we now show in Figure \ref{fig:relations_pops} the mass--size relation as panel (a) of Figure \ref{fig:relations_plot}, but now galaxies are colour coded by their stellar population properties (mean light-weighted stellar age, left panel; mean light-weighted metallicity, right panel). In this case, we now show the lines of iso-velocity dispersion (left to right, $\sigma=$20, 30, 50, 100, 200 and 250\, \kms) obtained directly from the Virial theorem ($M=5 \, \re \, \sigma^{2}/G$). For the comparison sample we are using the stellar populations published in \citet{Janz2016}, which are partly obtained from ATLAS$^\mathrm{3D}$ \citep{McDermid2015}. We also show CMGs from the previous plot (hexagons; \citet{Trujillo2009}; \citet{Ferre-Mateu2012,Ferre-Mateu2015,Ferre-Mateu2017}; \citet{Yildirim2017} and \citet{Spiniello2020}). But because we are now focusing on the stellar population properties, we further highlight those CMGs that are confirmed relics with crossed pentagons. The difference between the young CMGs and the relics is now very visible in the age panel, while it is hard to differentiate in the metallicity one.  We find that cEs in general follow the same properties as the more massive galaxies, rather than the stellar populations of the more extended galaxies of similar mass (dwarf ellipticals). In fact, they do not even seem to follow the lines of iso-velocity as suggested in \hyperlink{G+15}{G+15} but rather the lines of higher iso-velocity. Once the sample is enlarged (from the eight objects in \hyperlink{G+15}{G+15} to around two hundred here), the claimed trends disappear. This is dramatically apparent in the metallicity plot, where the bulk of cEs have higher metallicities than their counterparts in iso-velocity, and are more similar to galaxies of higher velocity dispersions, such as the CMGs. That is, cEs seem to follow a diagonal decreasing track to smaller sizes, masses and velocity dispersions (black arrow), which is a direct effect expected from the stripping process. We are thus suggesting that for many of these cEs that are the result of stripping, their progenitor galaxy could have been a CMG (Figure \ref{fig:cartoon}, channel $a$), and some other might even be the relic seed of $d$ channel in Figure \ref{fig:cartoon}. Such a connection between compact galaxies at all stellar masses has not been reported so clearly before and suggests that cE galaxies share a direct link to the inner-most parts of ETGs.

Another crucial property that can give us important clues on the origin of these objects is the abundance of alpha elements, [$\alpha$/Fe], which can be used as a global formation timescale clock in galaxies. Typically, large values of $\alpha$-enhancement are thought to be representative of very rapid formation timescales in ETGs. Early in the formation of the galaxy, only more massive stars evolve and explode as core collapse supernovae, producing large yields of alpha elements such as Mg and thus having large alpha-abundances. As time evolves, lower mass stars start also evolve and the explosions of the SNIa start to yield higher values of iron-based elements. This will naturally lower the overall [$\alpha$/Fe] (e.g.\ \citealt{Thomas1999}; \citealt{Ferreras2003}). This way, massive galaxies, which form early and fast, tend to have high alphas whereas low mass galaxies tend to exhibit lower alpha-enhancements due to their more extended SFHs (e.g.\ \citealt{Thomas2005}; \citealt{deLaRosa2011}; \citealt{McDermid2015}). 

\begin{figure*}
\centering
\includegraphics[width=1.02\textwidth]{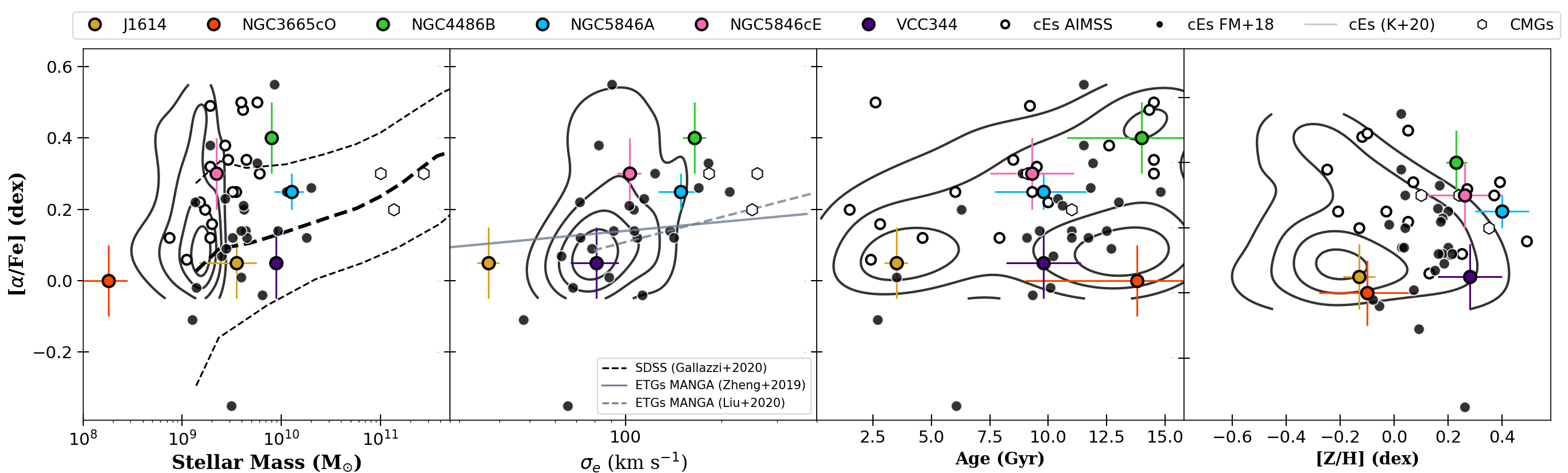}
\vspace{-0.2cm}
\caption{[$\alpha$/Fe] relations of cEs. This figure shows the relation of the $\alpha$-enhancement values obtained for our cEs (colour-coded following the previous figures), those from the AIMSS compilation studied in \citet{Janz2016} (open circles), FM+18 (black circles) and K+20 (density contours) with different structural and stellar population parameters. First and second panels show the relation with the mass of the galaxy (with the stellar mass and the velocity dispersion, respectively), including literature relations from SDSS \citep{Gallazzi2020} and MaNGA (\citealt{Zheng2019}; \citealt{Liu2020}). Third and fourth panels show the relation with the stellar mean age and metallicity (from the line-index approach).}
\label{fig:alpha_rels_plot}
\end{figure*}

Figure \ref{fig:alpha_rels_plot} shows different relations for the [$\alpha$/Fe]. The leftmost panel shows its relation with the stellar mass, driven by the recent relation found in \citet{Gallazzi2020} (dashed line). cEs show a large scatter around this relation, although the highest density countours of \hyperlink{K+20}{K+20} and some of our cEs are compatible with following the relation. Another proxy for the stellar mass is the velocity dispersion, so the second panel shows the [$\alpha$/Fe] -- $\sigma$ relation compared to the relation of \citet{Zheng2019} (solid grey line). We find that including the cEs in the scaling relations would actually make it steeper, similar to the slope that \citet{Liu2020} obtained when including only their more massive objects (grey dashed line). These are additional indications for the stripping nature, as most of these cEs have alpha-abundances larger than expected for their mass/sigma. The objects in our sample that seem to follow the trends are J1614, VCC~344 and NGC~3665cO (if we consider the instrumental resolution as an upper limit for the velocity dispersion). These three objects are the ones that have similar SFH shapes (Figure \ref{fig:sfhs_plot}). However, only J1614 is truly young, as seen in the third panel of Figure \ref{fig:alpha_rels_plot}, which shows the age--[$\alpha$/Fe] relation. While younger galaxies show relatively low values of alpha-enhancement, there is a large spread for old ones, flattening the relation. Nonetheless, the dichotomy between old/young cEs is still present. The last panel shows the relation of the alpha-elements with total metallicity. Here more enhanced galaxies seem to have lower metallicities. This was already hinted in \hyperlink{FM+18}{FM+18} but with this larger sample, the trend is now more clear. From this relations with the alpha abundance, we thus find that the results unequivocally reinforce the intrinsic origin for J1614 and the stripping one for NGC~4486B, NGC~5846A and NGC~5846cE.

One last property that can provide insight into the nature of cEs is the ratio within the dynamical and the stellar mass. It has been shown that the region covered by low-mass compact systems (cEs and UCDs), presents abnormally enhanced values (e.g.\ \citealt{Mieske2013}; \citealt{Forbes2014}; \citealt{Janz2015,Janz2016b}; \hyperlink{FM+18}{FM+18}). The true nature of this effect is still a matter of debate, as it can be accounted for through different processes. One possibility is the presence of dark matter (e.g.\ \citealt{Hasegan2007}; \citealt{Baumgardt2008}), while another possibility calls for having a different initial mass function. A third option is related to the presence of a black hole in the centre of the galaxy. In the case of intrinsic cEs, one would expect them to follow the local scaling relations governing low-mass galaxies, and therefore could host the so-long sought intermediate black holes (\citealt{Mezcua2017}; Ferr\'e-Mateu et al. in prep). But those resulting from the stripping of a larger, more massive galaxy, will instead host a super massive black hole (SMBH), as expected according to the original mass of the progenitor galaxy. This way, they will have a larger dynamical mass than their stellar mass \citep{Seth2014}. In addition, while the more massive progenitor is losing its stellar mass and shrinks due to the stripping or tidal effects, its central velocity dispersion remains almost unaltered, further emphasizing the deviations from the 1:1 relation \citep{Pfeffer2013}. Unfortunately, the search for such intermediate black holes and even the more massive SMBHs in the centers of cEs is a very challenging and difficult endeavour due to the small sizes of the objects, with only a handful of them reported to date (e.g.\ \citealt{Kormendy1997}; \citealt{vanderMarel1997}; \citealt{Mieske2013}; \citealt{Paudel2016}).

\begin{figure}
\centering
\includegraphics[scale=0.48]{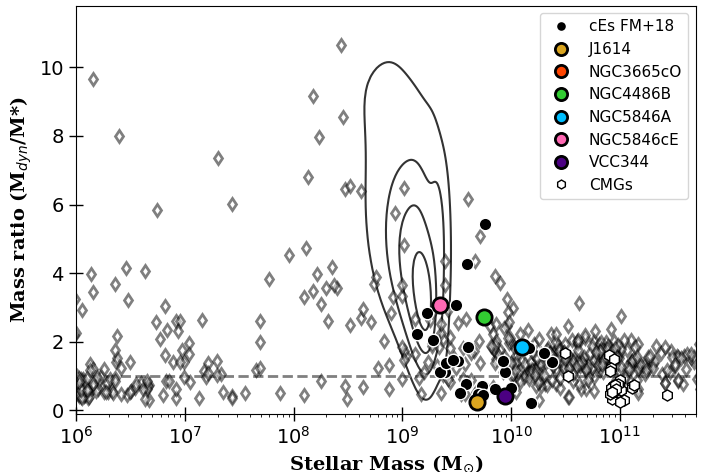}
\vspace{-0.2cm}
\caption{Dynamical-to-stellar mass ratio with galaxy mass. The relation between the dynamically inferred mass ($\mathrm{M_{dyn}}$) and that from the stellar populations as function of the stellar mass of different galaxies, with the cEs for which this information is available in the literature (those in FM+18) shown as solid black circles, the K+20 cEs as contours and this work cEs in coloured dots. The open diamonds correspond to galaxies from the AIMSS compilation. }
\label{fig:relations_massdyn}
\end{figure}

Therefore, studying the deviations in the $M_\mathrm{dyn}/M_{*}$ relation can provide an indirect proof for the (possible) existence of such SMBHs. Although we showed in \hyperlink{FM+18}{FM+18} that this was not the most determining property, it did help to shed light onto the origin of those cEs that extremely deviated from the 1:1 relation (e.g. $M_\mathrm{dyn}/M_{*}> 2$). The dynamical masses have been obtained from the Virial theorem for pressure-dominated systems: \\
\begin{equation}
M_\mathrm{dyn}\,=\,C\,G^{-1}\,\sigma^{2}\, \re, 
\end{equation}
\noindent
where $\sigma$ is the velocity dispersion, $\re$ is the effective radius of the system and \textit{C} is a virial coefficient. The latter value is given by the S\'ersic index of the system \citep{Bertin2002}, which is $C=6.5$ for UCDs and other compact systems such as cEs (\citealt{Mieske2013}; \citealt{Forbes2014}; \hyperlink{FM+18}{FM+18}). We have also computed these dynamical masses for the sample of \hyperlink{K+20}{K+20}, as shown by the density contours in Figure \ref{fig:relations_massdyn}. This figure also shows the values used in \hyperlink{FM+18}{FM+18} (black circles). NGC~5846cE, NGC~4486B and NGC~5846A are, by this order, the ones deviating the most from the relation, while J1614 and VCC~344 are closer to unity.

We suggest that, together with the rest of properties seen previously -- high [$\alpha$/Fe], old ages, high metallicities, and relatively high velocity dispersions/stellar masses, and being fast rotators -- NGC~5846cE, NGC~4486B and NGC~5846A undoubtedly share a stripped origin. Their bursty SFHs suggest that their progenitors had some reservoirs of gas, so it could have been either a spiral galaxy or a gas-rich ETG. Given the radial gradients in age and metallicity, the latter seems to be a more plausible progenitor. Their local and global environment (all three have an associated host and belong to groups of galaxies) further supports this claim. We then confirm J1614 as a low-mass, intrinsic cE due to its extended SFH, very young mean age, low metallicity, low alpha-abundance, very low velocity dispersion and because it follows the majority of the scaling relations of massive ETGs. This is the only cE that has no associated host in our sample and although it belongs to a cluster, it is located in the outskirts of it. This is where these type of low mass galaxies with extended SFHs tend to be found (e.g.\ \citealt{Thomas2005}; \citealt{Ferre-Mateu2014}). For VCC~344 and NGC~3665cO it is not as straightforward as for the other cEs in our sample. They both have similar SFHs to ETGs of \mstar$\sim 10^{10-11}$\msun (e.g. \citealt{McDermid2015}) but both follow the [$\alpha$/Fe] relations. VCC~344 is close to unity in Figure \ref{fig:relations_massdyn} and its stellar population gradients are very similar to those of low-mass ETGs. However, they are both extreme outliers in the mass--metallicity relation. Altogether and considering that they are both located in dense environments (one is in a cluster, the other in a group), we suggest that they are more compatible with being the result of stripping. But in both cases, the host galaxy would have been a massive, gas-poor elliptical galaxy (either compact or extended).

\vspace{-0.5cm}
\section{Conclusions}\label{section:conc}

cEs are a very rare family of galaxies that populate the low-mass/luminosity end of the realm of ETGs. Here we have obtained and analyzed new IFU data from KCWI at the Keck II Telescope for a sample of 6 nearby cEs. Of these, only NGC~4486B was previously studied with an IFU, while NGC~5846A, VCC~344 and J1614 only had long-slit spectroscopy. The remaining two cEs are newly reported galaxies and not much spectral information has been gathered to date about them (NGC~5846cE and NGC~3665cO). 

We measured the first order kinematic and rotation parameters, as well as the stellar population properties, i.e.\ stellar ages, total metallicities, alpha enhancements and star formation histories (SFHs). We compared our sample with a larger sample of cEs and with samples of ETGs, LTGs and compact massive galaxies. Despite the caveat of the different physical scales that set limitations on our power to compare them, this can help in addressing the possible progenitors of cE in the case of stripped cEs. Our results are here summarized:

\begin{itemize}
    \item All our cEs have similar structural parameters, with \ssc{} indices $n \la 1$ and high axial ratios ($b/a \ga 0.7$). Their surface brightnesses at 1$\re$ are around $\mu_{V}\sim$ 19--20 mag\,arcsec$^{-2}$. They have a range of sizes, from the most compact NGC~3665cO (136\,pc) to the largest NGC~5846A ($\re$=\,520\,pc). 
    \item We find that cEs have moderate rotational velocities but they reveal a wide range of velocity dispersions ranging from 36--170 \kms. Their $\lambda_R$ values generally place them as (core-less) fast-rotators, due to their moderate rotation for their low ellipticites. Their kinematic trends can not provide definitive clues on the type of progenitor and we need the information from the stellar populations.
    \item Our cEs show mostly uniformly old ages ($\sim$12\,Gyr) with the bulk of their stars formed very early on. They show mild decreasing age gradients (of $\sim$2\,Gyr within 1~$\re$) but steep metallicity gradients with high central metallicities ([$Z$/H]$\sim$0.1--0.2\,dex). We have compared the shapes of these gradients with literature ones (from the MASSIVE, MaNGA and CALIFA surveys) finding that the majority of the cEs follow similar trends to ETGs, in particular to the low-mass ones. Only J1614 has strikingly different gradients. Its SFH is also different, being younger ($\sim$7\,Gyr) and more extended. It is also metal-poor ([$Z$/H]$\sim -$0.15\,dex) than the other cEs. All cEs show almost constant values of [$\alpha$/Fe] throughout their structures, with NGC~4486B, NGC~5846A and NGC~5846cE being alpha-enhanced and J1614, NGC~3665cO and VCC~344 having solar abundances. 
    \item The majority of the cEs are outliers in the mass--metallicity relation, which may be interpreted as a signature for a stripping origin. We can confirm that NGC~4486B, NGC~5846A and NGC~5846cE are, indeed, cEs that were formed following this formation scenario. This is reinforced by the fact that they are also outliers in the relations of [$\alpha$/Fe] and $\mathrm{M_{dyn}/M_{*}}$. Their SFHs show several star forming episodes, which would suggest a progenitor that was gas-rich and the stripping process involved gas flows. In the case of NGC~5846A the progenitor is more likely to be of the spiral-type due to this galaxy rotational parameters. NGC~3665cO and VCC~344 follow the [$\alpha$/Fe] and $\mathrm{M_{dyn}/M_{*}}$ but are outliers in the mass--metallicity relation. Moreover, they have SFHs that are typically seen in more massive ETGs or compact massive galaxies, which would suggest that these galaxies are the result of stripping such types of galaxies. Only J1614 is unequivocally an in-situ-formed cE. It follows all the scaling relations studied and shows an extended SFH, with younger mean ages and is more metal-poor than average. 
    \item We find that cEs have the same stellar populations of their massive compact counterparts of higher velocity dispersion. In the mass--size diagram, objects under stripping will follow diagonal tracks whereby they have their size, stellar mass and $\sigma$ reduced. We therefore caution that one should be comparing this rare family of galaxies to such massive compact galaxies or at most, to the innermost regions of large ETGs or LTGs. Unfortunately, to date the majority of surveys that provides 2D maps and trends do not have the power to resolve such inner regions or do not provide the results in a physical scale, making these comparisons difficult. 
\end{itemize}

Although cEs are rare, it is clear that they can form via a mixture of different channels, as we have presented in Figure \ref{fig:cartoon}. Unfortunately, whether there is a preferred channel to form this galaxies and whether it is mass dependent, is still an open question due to the selection and observational biases discussed throughout this paper. Future studies should aim to obtain high quality and spatially resolved spectra, such as the one presented here or in \hyperlink{FM+18}{FM+18}, for large samples of cEs in order to estimate the relative frequency of each channel. The fact that some cEs are formed in-situ and are hence intrinsically low-mass compact galaxies highlights them as targets to search for the elusive intermediate black holes. Furthermore, the fact that a connection to the more massive but compact galaxies is seen for the first time, suggests that the realm of compact galaxies might follow its own formation path than the more extended ETGs. 

\section*{Data Availability}
Data are available from the Keck Observatory Archive at
https://www2.keck.hawaii.edu/koa/public/koa.php

\section*{Acknowledgements}
We thank Alex Colebaugh for key contributions with the cE sample selection. We also thank Suk Kim and Soo-Chang Rey for providing the list of cEs published in their recent paper. We also thank the referee for the suggestions that improved the quality of the paper. AFM has received financial support through the Postdoctoral Junior Leader Fellowship Programme from `La Caixa' Banking Foundation (LCF/BQ/LI18/11630007). AFM, DAF and RM thank the ARC for financial support via DP160101608. AJR was supported by National Science Foundation grant AST-1616710 and as a Research Corporation for Science Advancement Cottrell Scholar. AA was supported in part by NASA program HST-GO-14747, contract NNG16PJ25C, and grant 80NSSC18K0563, and NSF award 1828315. RMcD is the recipient of an Australian Research Council Future Fellowship (project number FT150100333). \\

This work was supported by a NASA Keck PI Data Award, administered by the NASA Exoplanet Science Institute. Data presented herein were obtained at the W. M. Keck Observatory, in part from telescope time allocated to the National Aeronautics and Space Administration through the agency's scientific partnership with the California Institute of Technology and the University of California. The Observatory was made possible by the generous financial support of the W. M. Keck Foundation. The authors wish to recognise and acknowledge the very significant cultural role and reverence that the summit of Maunakea has always had within the indigenous Hawaiian community.  We are most fortunate to have the opportunity to conduct observations from this mountain. \textit{M\=alama ka '\=aina}.

\bibliographystyle{mnras}
\bibliography{kcwi_cEs}



\bsp	
\label{lastpage}
\end{document}